\documentclass[]{jfm}

\usepackage{graphicx}
\usepackage{newtxtext}
\usepackage{newtxmath}
\usepackage{array}
\usepackage{natbib}
\usepackage{hyperref}
\hypersetup{
    colorlinks = true,
    urlcolor   = blue,
    citecolor  = black,
}

\newcommand{\RomanNumeralCaps}[1]
\linenumbers

\title{Instability and breaking of internal waves in a horizontal shear layer}

\author{Sam F. Lewin\aff{1}
  \corresp{\email{slewin@berkeley.edu}},
 Alexis K. Kaminski\aff{1},
 Arun Balakrishna\aff{2}
 \and Miles M.P. Couchman\aff{3}}

\affiliation{\aff{1}Department of Mechanical Engineering, University of California, Berkeley, CA 94720, USA. 
\aff{2} Center for Turbulence Research, Stanford University, Stanford, CA 94305, USA. 
\aff{3} Department of Mathematics and Statistics, York University, Toronto, ON M3J 1P3, Canada.}

\begin{document}
\maketitle

\begin{abstract}
The behaviour of internal waves propagating in a background shear flow is studied in the case where the direction of shear is orthogonal to gravity. Ray-tracing theory is used to predict properties of the wave state at locations where instability occurs. Local wave energy growth is found to result from two distinct mechanisms: an increase in wave steepness due to refraction by the shear, or an increase in streamwise velocity perturbations due to wave advection of the background flow. Based on the initial conditions, a dimensionless perturbation energy ratio $F$ is constructed to predict the relative importance of these two mechanisms in facilitating wave-breaking. When $F$ is small and waves become locally steep, perturbation kinetic and potential energy remain approximately equipartitioned and subsequent instabilities are expected to develop due to a combination of shear and convection. On the other hand, as $F$ increases, kinetic energy dominates and wave advection of momentum may instead cause breaking to become increasingly driven by enhanced vertical shear. To test these predictions, fully nonlinear direct numerical simulations are conducted, spanning a range of wave-breaking dynamics. Good qualitative agreement with the theory is found despite substantial departures from the underlying assumptions. Wave breaking leads to significant turbulent dissipation, which in some cases greatly exceeds the initial wave energy. Momentum and energy transfers between the wave, background flow and turbulence are found to be sensitive to the dynamics of breaking, as are the mixing properties.
\end{abstract}

% Notably, the latter mechanism can lead to much enhanced kinetic energy whilst the potential energy remains essentially unchanged, whereas equipartition is maintained for the former


% {\bf MSC Codes }  {\it(Optional)} Please enter your MSC Codes here

\section{Introduction}
In both the atmosphere and the ocean, a significant portion of energy is stored in and transported by internal gravity wave motions. The means by which these waves break and dissipate their energy via small-scale turbulence is a classical and important problem with an extremely broad scope, encompassing a variety of linear and nonlinear interactions and instabilities over a range of scales. Wave breaking is often inferred observationally by considering one-dimensional vertical spectra of shear and isopycnal strain. At vertical scales larger than O(10m), there is generally a close match between observed spectra and the empirical Garrett-Munk spectrum modelling dynamics dominated by weakly nonlinear internal waves \citep{garrett1975space}. On the other hand, energetic turbulence typically exhibits an inertial subrange over scales smaller than the Ozmidov length scale (below which the vertical extents of turbulent motions are not affected at leading order by the stratification). It is routinely observed that the smallest scales of internal waves are larger than the largest scales of isotropic turbulence \citep{kunze2019unified,dasaro2022internal}. The dynamics across the resulting intermediate range of scales are in general not well understood, though evidence suggests they may be of leading order importance in determining the properties of turbulence and mixing at dissipative scales. For example, it is becoming increasingly apparent from observations in a variety of environments that the mixing efficiency of stratified turbulence -- often assumed to be constant throughout the ocean interior -- exhibits up to an order of magnitude variability over spatial and temporal scales significantly larger than those associated with turbulence (for recent examples, see \citet{alford2025buoyancy,lewin2025multiscale,ijichi2018observed}). The implication is therefore that the dynamics at the scales of wave breaking are of leading order importance for the subsequent properties of mixing, a feature framed by \citet{caulfield2021layering} in the terms `history matters'. 

To capture the influence of dynamics at scales over which internal waves dominate energetics and to model downscale energy transfer in a mathematically tractable way, our study will consider the evolution and eventual instability of linear monochromatic internal waves under the Boussinesq approximation and in the limit of weak rotation $f\ll N$, where $f$ and $N$ are the Coriolis and buoyancy frequency, respectively. Floquet theory may be used to show that even plane periodic waves of this kind are unstable at arbitrarily small amplitudes to a parametric instability driven by wave self-interaction \citep{mied1976occurrence,klostermeyer1982parametric,lombard1996instability}. However, the growth rate of this instability is typically expected to be small, in particular compared with the rate of modulation of the local wave amplitude due to wave-mean flow interactions (see, e.g. \citealp{sutherland2006internal}). In either case, when the wave amplitude is sufficiently large (such that instabilities grow quickly) or the frequency is sufficiently small, a local linear stability analysis may be performed on vertical velocity and buoyancy profiles derived from or approximating the wave polarisation relations \citep{fritts2003gravity,dunkerton1997shear}. This provides potentially useful information about, for example, the roles of convection and shear in the resulting turbulent state \citep{lelong1998inertiaa,lelong1998inertiab}, which are found to vary significantly with the properties of the background wave. In this framework, there is therefore motivation to explore the range of possible unstable configurations that are dynamically accessible during the lifecycle of an internal wave. 

A classical paradigm for the modulation and eventual breaking of small-scale, vertically propagating wave packets is the rapid increase in vertical wavenumber due to strong refraction near critical levels in a vertical shear flow. Such critical levels, originally studied by \citet{booker1967critical} and \citet{bretherton1966propagation}, are locations where the horizontal phase velocity of a wave group approaches the speed of the horizontal background flow. A Wentzel-Kramers-Brillouin-Jeffreys (WKBJ) analysis may be used to demonstrate that the intrinsic frequency (the frequency measured by an observer travelling at the speed of the background flow) decreases and approaches zero at the critical level (in the absence of rotation). The wave becomes increasingly steep leading to the possibility of breaking, as has been captured experimentally by \citet{koop1981preliminary, thorpe1981experimental} and in turbulence-resolving numerical simulations by \citet{winters1994three, dornbrack1998turbulent,howland2021shear}. Approximate local forms for the vertical velocity and density profiles indicate that strong vertical shear and statically unstable density profiles are both possible \citep{winters1992instability}; correspondingly, the emerging modes of instability are dynamically rich and inherently three-dimensional, with important consequences for the resulting turbulent energetics \citep{winters1994three}. A notable success of this paradigm has been the eikonal theory of ray tracing, which, when applied to test waves in a measured background shear flow, may be used to estimate the fraction of energy that is cascaded to small scales and eventually dissipated \citep{henyey1986energy}. This method forms the basis of the finescale parameterisation approach for estimating energy dissipation from measurements of the larger-scale internal wave field \citep{polzin2014finescale}. 

The problem of internal wave modulation by a \emph{horizontal} shear has received relatively less attention in the literature. In this case, WKBJ analysis indicates the existence of a new type of critical level where the intrinsic wave frequency approaches the buoyancy frequency \citep{olbers1981propagation,basovich1984internal, badulin1985trapping}, resulting in dynamics that are characteristically distinct from the vertical shear scenario described above. These critical levels are sometimes referred to as baroclinic critical layers and have been shown to play a significant role in the dynamics of rotating stratified shear flows \citep{marcus2013three} and tilted stratified vortices \citep{boulanger2007structure}. Nonlinear dynamics within baroclinic critical levels under a steady wave forcing in a stratified uniform horizontal shear flow have been described by \citet{wang2020nonlineari,wang2020nonlinearii}. More relevant to this study is the work of \citet{staquet2002transport}, who investigated the transient problem of an inertia-gravity wave packet approaching a critical level in a horizontal shear layer using numerical simulations. Wave breaking and significant subsequent turbulence production was observed due to a transient growth in wave energy extracted from the shear in the vicinity of the critical level. Building on this result, \citet{bakas2009gravitya,bakas2009gravityb} analysed the process by which linear internal waves may extract energy from a background unbounded uniform horizontal shear flow, thus potentially leading to instability and wave breaking. They showed that, in fact, two distinct mechanisms for energy perturbation growth exist. The first corresponds to wave amplitude growth due to extraction of energy from the background shear through downgradient Reynolds stresses, consistent with the dynamics near critical levels observed by \citet{staquet2002transport}. The second results from wave advection of the mean flow momentum that may produce significant vertical shears, thus providing an alternative route to wave breaking. This latter mechanism operates in the same way as the well-known lift-up mechanism for unstratified flows \citep{ellingsen1975stability}.

Motivated by the discussion above, in this study we revisit the initial value problem considered by \citet{staquet2002transport} consisting of a monochromatic internal wave packet in a background  horizontal shear layer
\begin{equation}\label{eq:backgroundflow_dimensional}
    \mathbf{U}^*(y^*) = \Delta u^* \tanh(y^*/h^*)\mathbf{e}_x,
\end{equation}
where $\Delta u^*$ and $h^*$ are constants characterising the velocity and length scales of the shear and stars denote dimensional variables. The linear stability of this base flow in a uniform stratification was studied by \citet{deloncle2007three}, who showed that the most unstable mode is two-dimensional with zero vertical wavenumber. However, as the (horizontal) Froude number $Fr_h = \Delta u^* /N^*h^*$ decreases, the growth rates of modes with vertical wavenumber $k_z>0$ increase, thereby increasing the tendency for vertical structure to emerge in practice. Indeed, the fully nonlinear simulations of \citet{basak2006dynamics} highlight such three-dimensional behaviour, though turbulence was not sustained for Froude numbers $Fr_h\leq 1$. Recently, \citet{lewin2024evidence} demonstrated the existence of energetic turbulence in this strongly stratified regime, provided an additional small-amplitude wave-like perturbation was supplied as an initial condition. Waves were able to rapidly extract energy from the horizontal shear and amplify vertical shear in the flow by multiple orders of magnitude through the lift-up mechanism described by \citet{bakas2009gravitya}, resulting in a significant departure from the dynamics observed by \citet{basak2006dynamics}. In this study, the linear instability of the stratified horizontal shear flow is suppressed by choosing a sufficiently large shear layer width $h^*$ relative to the domain size in the $x$-direction (precisely, unstable perturbations are restricted to wavelengths $\lambda_x^*>2\pi h^*$ \citep{deloncle2007three}).

We investigate the local structure and amplitude of wave-like perturbations in the shear flow \eqref{eq:backgroundflow_dimensional} using ray-tracing theory derived under the WKBJ approximation. A principle aim is to explore whether this method, similar to that of \citet{winters1992instability} for the vertical shear case, can make useful predictions regarding the local flow fields of the waves from which instabilities may develop and lead to breaking. In particular, we wish to determine whether the effects of the two mechanisms for wave energy growth in a horizontal shear described by \citet{bakas2009gravitya} may be captured by the local theory. Using this theory as a guide, we then proceed to construct a parameter space (based on the scales associated with the wave and shear flow) for investigation using fully nonlinear direct numerical simulations (DNS). In cases where wave-breaking leads to turbulence, a key question to be addressed is: how do the eventual bulk momentum and energy exchanges between the mean flow and perturbations depend on the route to wave breaking? An obvious point of reference is the vertical critical level problem, where it has been typically observed that a significant fraction of the incoming wave energy is absorbed by the mean flow near the critical level, even in cases where wave-breaking results in considerable turbulent dissipation \citep{howland2021shear,winters1994three, thorpe1981experimental}. 

The remainder of the manuscript proceeds as follows. In \S\ref{sec:setup}, we describe the basic problem setup and detail the ray-tracing theory used in the analysis. In \S\ref{sec:simulations}, we outline a suite of DNS with parameters chosen based on predictions from the theoretical analysis. A qualitative overview of the DNS results and a comparison with the theory is presented in \S\ref{sec:results}, followed by a detailed description of momentum and energy budgets in \S\ref{sec:momentum} and \S\ref{sec:energy}. Finally, we discuss some broader implications of the results and conclude in \S\ref{sec:conclusions}.

\section{Problem setup and ray-tracing theory}\label{sec:setup}
\subsection{Initial value problem}\label{sec:ivp}
We consider a Boussinesq fluid with uniform density stratification characterised by buoyancy frequency $N^{*2}$. Precisely, the buoyancy field is decomposed as $b^* = N^{*2}z^* + \theta^*$, where $\theta^* = -g^*\rho^*/\rho_0^*$ is the buoyancy perturbation corresponding to density perturbation $\rho^*$ away from hydrostatic balance. Here, $g^*$ is the acceleration due to gravity in the vertical ($z$) direction and $\rho_0^*$ is a constant reference density. The equations of motion are non-dimensionalised with time and length scales obtained from the buoyancy frequency $N^*$ and a characteristic wavenumber (or inverse length scale) $K^*$ to be associated with the initial wave. In this way, dimensionless variables are defined as 
\begin{equation}
    t = N^*t^*, \quad x_i = K^* x_i^*, \quad u_i = \frac{K^*}{N^*} u_i^*, \quad
    \theta = \frac{K^*}{N^{*2}}\theta^*, \quad 
    p = \frac{K^{*2}}{N^{*2}} \frac{p^*}{\rho_0^*},
\end{equation}

so that the equations of motion are 
\begin{subequations}
\begin{eqnarray}
%write D/Dt as partial plus advective derivative
    &\displaystyle \frac{\mathrm{\partial}\mathbf{u}}{\mathrm{\partial}t} + \mathbf{u}\cdot\nabla \mathbf{u} = -\nabla p + \theta\mathbf{e}_z+ \frac{1}{Re_w} \nabla^2\mathbf{u}, \label{eq:ns1} & \\
    &\displaystyle \nabla \cdot \mathbf{u} = 0, \label{eq:ns2} & \\
    &\displaystyle \frac{\mathrm{\partial}\theta}{\mathrm{\partial}t} + \mathbf{u}\cdot \nabla \theta + w = \frac{1}{Re_wPr}\nabla^2 \theta. \label{eq:ns3} &
\end{eqnarray}
\end{subequations}
The wave Reynolds number $Re_w \equiv N^*/(K^{*2}\nu^*)$ (cf. \citealp{fringer2003dynamics}) and Prandtl number $Pr \equiv \nu^*/\kappa^*$ are defined in terms of the molecular kinematic viscosity $\nu^*$ and buoyancy diffusivity $\kappa^*$. Note that, under this choice of non-dimensionalisation, the coefficient in front of the buoyancy term in the momentum equation \eqref{eq:ns1} is unity. 

In the absence of any background flow, monochromatic periodic linear internal waves with wave number $\mathbf{k}_0 = (k_{x0},k_{y0},k_{z0})$ and frequency $\Omega_0$ are given by 
\begin{eqnarray}\label{eq:wkbwave}
    [u,v,w,\theta] = [U_0,V_0,W_0,\Theta_0]s_0\exp(i\mathbf{k}_0\cdot \mathbf{x} - i\Omega_0 t),
\end{eqnarray}
where the amplitude coefficients are defined by the polarisation relations 
\begin{equation}\label{eq:polarisation}\begin{split}
    \Theta_0 = \frac{1}{k_{z0}} ,\quad U_0 = -\frac{\mathrm{i} k_{x0} }{\Omega_0 |\mathbf{k}_0|^2}, \\
    V_0 = -\frac{\mathrm{i} k_{y0}}{\Omega_0|\mathbf{k}_0|^2}, \quad W_0 = \mathrm{i}\frac{\Omega_0}{k_{z0}},
    \end{split}
\end{equation}
and the frequency $\Omega_0$ is given by the dispersion relation \begin{eqnarray}\label{eq:dispersionrelation}
    \Omega_0^2 = \frac{k_{x0}^2+k_{y0}^2}{|\mathbf{k}|^2}.
\end{eqnarray}
The wave steepness $s_0$ appearing in \eqref{eq:wkbwave} is a dimensionless amplitude defined such that $s>1$ corresponds to the existence of regions of local static instability in the total buoyancy field with $\partial b/\partial z <0$. 

\begin{figure}
\centering
\includegraphics[width=0.8\textwidth]{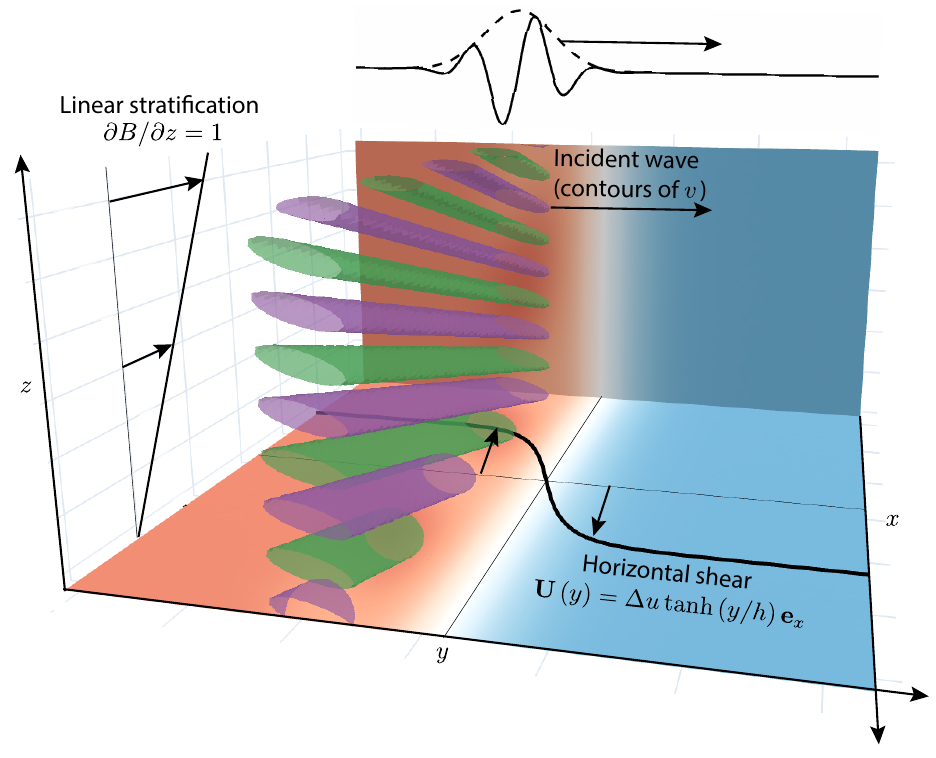}% Here is how to import EPS art
\caption{\label{fig:1} Schematic showing a representative initial condition for the problem under consideration. The shape and location of the incident wave packet is shown by plotting isocontours of the $y$-component of the velocity field ($v$). Notably, the wave packet is attenuated in $y$ based on the exponential envelope defined in (\ref{eq:wavepacket}). The tilt of the isocontours in the $xz$ plane is governed by the wavenumber ratio $k_x/k_z$.}
\end{figure}
We will consider a wave packet bounded by a Gaussian envelope in $y$ of width $a$, centered at an initial location $y_0$. At time $t=0$, the velocity and buoyancy perturbation fields are given by $\mathbf{u}_\mathrm{wave}(\mathbf{x})$ and $\theta_\mathrm{wave}(\mathbf{x})$, where 
\begin{eqnarray}\label{eq:wavepacket}
    [u_\mathrm{wave},v_\mathrm{wave},w_\mathrm{wave},\theta_\mathrm{wave}] = [U_0,V_0,W_0,\Theta_0]s_0\exp\left(\mathrm{i}\mathbf{k}_0\cdot \mathbf{x}-\frac{(y-y_0)^2}{a^2}\right),
\end{eqnarray}
with amplitudes given by \eqref{eq:polarisation}. Taking $y_0<0$, the sign of $k_{y0}$ is chosen to be positive such that the wave packet propagates in the positive $y$-direction. We note that wave packets of this form have also been studied by both \citet{staquet2002transport} and \citet{bakas2009gravitya}. In dimensionless form, the background shear flow into which the wave will propagate is 
\begin{equation}\label{eq:backgroundflow}
    \mathbf{U}(y) = \Delta u \tanh(y/h)\mathbf{e}_x,
\end{equation}
where $\Delta u = \Delta u^* K^*/N^*$ and $h = h^*K^*$ are dimensionless constants. Putting everything together, the initial velocity and buoyancy fields are $\mathbf{u}(\mathbf{x},t=0) = \mathbf{U}(y) + \mathbf{u}_\mathrm{wave}(\mathbf{x})$ and $ \theta(\mathbf{x},t=0) = \theta_\mathrm{wave}(\mathbf{x})$. A schematic of the flow described by this initial condition is shown in figure \ref{fig:1}. 

\subsection{Linear dynamics}
% Though the dynamics we will eventually consider are fully nonlinear, qualitative features of the wave-shear interaction emerge from a linear analysis. This allows us to construct a parameter space for direct numerical simulations within which a rich variety of dynamics are observed.

We wish to describe the propagation of the wave packet given by \eqref{eq:wavepacket} in the background flow \eqref{eq:backgroundflow}, with an eye towards predicting properties of the local wave state that determine the dynamics of wave breaking. A natural approach is to start from the inviscid equations of motion linearised about the mean shear $U(y)$
\begin{subequations}\label{eq:linearized}
\begin{eqnarray}
    &\displaystyle \frac{\partial u}{\partial t} + U(y)\frac{\partial u}{\partial x} + v\frac{\partial U}{\partial y} = -\frac{\partial p}{\partial x},& \label{eq:u}\\
    &\displaystyle \frac{\partial v}{\partial t} + U(y)\frac{\partial v}{\partial x} = -\frac{\partial p}{\partial y},& \label{eq:v}\\
    &\displaystyle \frac{\partial w}{\partial t} + U(y)\frac{\partial w}{\partial x} - \theta = -\frac{\partial p}{\partial z},& \label{eq:w}\\
    &\displaystyle \frac{\partial \theta}{\partial t} + U(y)\frac{\partial \theta}{\partial x} +w= 0,& \label{eq:theta}
\end{eqnarray}
\end{subequations}
seeking WKBJ solutions of the form
\begin{eqnarray}\label{eq:wkbsolution}
    [u,v,w,\theta, p] = [\tilde{U}, \tilde{V}, \tilde{W}, \tilde{\Theta}, \tilde{P}](\mathbf{x},t)s(\mathbf{x},t) \exp(\mathrm{i}\varphi(\mathbf{x},t)/\epsilon) + O(\epsilon),
\end{eqnarray}
where $\epsilon$ is a small parameter characterising the relative scale of phase variations. In the problem at hand, the WKBJ framework describes waves with wavelengths that are asymptotically small relative to the width of the wave envelope $a$ and the width of shear layer $h$. Precisely, a local frequency $\omega(\mathbf{x},t)$ and wavenumber $\mathbf{k}(\mathbf{x},t)$ may be defined in terms of the phase function $\varphi(\mathbf{x},t)$ as
\begin{eqnarray}
    \omega = -\frac{1}{\epsilon}\partial_t \varphi, \quad \mathbf{k} = \frac{1}{\epsilon}\nabla \varphi.
\end{eqnarray}

Substituting into \eqref{eq:linearized} above, one obtains locally a set of algebraic equations for $U_0$, $V_0$, $W_0$, $\Theta_0$ and $P_0$. Eliminating variables, at leading order the incompressibility condition gives
\begin{eqnarray}\label{eq:wkb_dispersion}
    |\mathbf{k}|^2\Omega^3 - |\mathbf{k}_h|^2\Omega -\mathrm{i}\frac{\partial U}{\partial y}k_xk_y(\Omega^2-1)    = 0,
\end{eqnarray}
where $\Omega \equiv \omega - Uk_x$ is the intrinsic frequency measured by an observer travelling at the speed of the background flow. The frequency $\omega$ measured by a stationary observer is often referred to as the absolute (or extrinsic) frequency. The two are related by the Doppler shift term $U(y)k_x$. For a nontrivial balance in \eqref{eq:wkb_dispersion}, we must have $\Omega = O(1)$. It follows that the first two terms are $O(\epsilon^{-2})$, so that the standard dispersion relation \eqref{eq:dispersionrelation} is recovered when $(\partial U/\partial y )k_xk_y = O(\epsilon^{-1})$.  Because $k_x$ is \textit{a priori} assumed $O(1/\epsilon)$, the usual approach in the literature is to assume that $U$ (and hence $\partial U/\partial y$) is $O(\epsilon)$ (see, e.g. \citealp{olbers1981propagation}). However, it will be shown below that, due to the fact that the background flow does not vary in the $x$ direction, $k_x$ is constant along rays. Therefore, the assumption that $k_x=1/\epsilon$ and hence $U = O(\epsilon)$ may be relaxed (provided the wave amplitude varies sufficiently slowly in $x$), as \citet{bretherton1966propagation} himself pointed out. Thus we may instead consider the case $k_x = O(1)$ with $U=O(1)$, in which case local polarisation relations are given by
\begin{equation}\label{eq:wave}
\begin{array}{>{\displaystyle}c >{\displaystyle}c}
\tilde{\Theta} = \frac{1}{k_{z}}, 
& \tilde{U} = -\frac{\mathrm{i}  k_{x}}{\Omega |\mathbf{k}|^2} 
        - \frac{k_y}{\Omega^2 |\mathbf{k}|^2} \frac{\partial U}{\partial y},  \\[0.5em]
\tilde{V} = -\frac{\mathrm{i}  k_{y}}{\Omega |\mathbf{k}|^2} , 
& \tilde{W} = \mathrm{i} \frac{ \Omega}{k_{z}}.
\end{array}
\end{equation}
The existence of the second term in the expression for $\tilde{U}$ in \eqref{eq:wave} at leading order whilst the local dispersion relation remains equivalent to \eqref{eq:dispersionrelation} is novel to the horizontally sheared system in the case when $U=O(1)$. It is important to keep in mind here that $s$, $\mathbf{k}$ and $\Omega $ are functions of $\mathbf{x}$ and t.
Changes in $\mathbf{k}$ (and hence $\Omega$, via the dispersion relation) may be determined using the standard kinematic theory of ray tracing, which involves computing wave properties along ray paths $\mathbf{x}(t)$ that are everywhere tangent to the group velocity field measured by a stationary observer $\mathbf{c}_g = \partial \omega/ \partial\mathbf{k} = U\mathbf{e}_x  + \partial \Omega/\partial \mathbf{k}$. Precisely,
\begin{eqnarray}\label{eq:wkbraypaths}
    \frac{d x}{d t} = U + \frac{\partial \Omega}{\partial k_x}, \ \ \frac{d y}{d t} = \frac{\partial \Omega}{\partial k_y},  \ \ \frac{d z}{d t} = \frac{\partial \Omega}{\partial k_z},
\end{eqnarray}
where $d/dt = \partial/\partial t + \mathbf{c}_g\cdot \nabla$. The change in $\mathbf{k}$ along rays is given by
\begin{eqnarray}
    \frac{d k_x}{dt} = 0 , \ \ \frac{d k_y}{dt} = -k_x\frac{\partial U}{\partial y}, \ \  \frac{d k_z}{dt} = 0.\label{eq:wkbwavenumbers}
\end{eqnarray}
This justifies the relaxation of the assumption $k_x=O(1/\varepsilon)$ above. Note in particular that when $k_x=0$ initially such that the group and phase velocities are orthogonal to the direction of the background flow, $k_y$ and hence $\Omega$, remain constant, that is, the intrinsic wave properties are entirely unaffected by the shear.  
% It is expected that changes in the local wavenumber $k_y$ occur on the slow timescale, so that $k_x\partial U/\partial y =O(1)$. This then implies the third term in the brackets in the dispersion relation is $O(\varepsilon)$ and can indeed be neglected at leading order. 

\subsubsection{Trapping and turning levels}
Given $U(y)=\Delta u \tanh(y/h)$, it follows from \eqref{eq:wkbwavenumbers} that $k_y$ increases or decreases monotonically for $k_x<0$ or $k_x>0$, respectively. We assume henceforth that $k_y>0$ initially, so that $k_x<0$ means $k_y>0$ for all times. In particular, the situation may arise where $k_y\to \infty$, in which case the intrinsic frequency approaches the  buoyancy frequency $\Omega \to 1$ (corresponding to $\Omega^*\to N^*$ in dimensional variables). The location $y=y_t$ where $\Omega = 1$ (when it exists) is referred to as a trapping level, where the group velocity in the $y$-direction approaches zero asymptotically. (Formally however, the WKBJ theory becomes invalid before the trapping level is reached.) It follows from the Hamiltonian structure of \eqref{eq:wkbraypaths} and \eqref{eq:wkbwavenumbers} that $\omega$ is constant along rays (e.g. \citealp{sutherland2010internal,buhler2014waves}). Then, equating the initial value of $\omega=\omega_0$ (for which the intrinsic frequency $\Omega = \Omega_0)$ with the value at the trapping level determines $y_t$ implicitly as
\begin{eqnarray}
    \tanh(y_t/h) = \frac{\Omega_0-1}{\Delta u k_x}+\tanh(y_0/h), \label{eq:yt_implicit}
\end{eqnarray}
where $y_0$ is the initial location of the wave packet. A detailed description of the asymptotic approach to $y_t$ is provided by \citet{badulin1985trapping}. The trapping level exists only when the right hand side of equation (\ref{eq:yt_implicit}) falls between $-1$ and $1$. Assuming without loss of generality that the wave packet propagates in the positive $y$ direction and is also located initially outside of the shear layer such that $\tanh(y_0/h)\approx -1$, it follows that we must have $0\leq (\Omega_0-1)/(\Delta u k_x)\leq 2$ for the trapping level to exist.

When $k_x>0$, $k_y$ decreases and internal reflection may occur where $k_y$ passes through $0$ and becomes negative. However, the WKBJ analysis fails for small $k_y$ and a separate solution must be obtained in the vicinity of the turning level or an alternative approach used (e.g. \citealp{bakas2009gravitya}). The location of the turning level where $k_y=0$ (when it exists) may be obtained from an equation analogous to (\ref{eq:yt_implicit}). 
% or, in dimensionless terms,
% \begin{eqnarray}
%     \frac{1}{\Omega}\frac{\partial U}{\partial y}\frac{k_x}{|\mathbf{k}|} = O(\varepsilon).
% \end{eqnarray}

\subsubsection{Kinetic and potential energy partition}
% In the WKB approximation, the velocities and buoyancy in the system are also assumed to be proportional to $s\exp(i\theta/\varepsilon)$, where the constants of proportionality are determined straightforwardly by substitution into the linear equations of motion. For example, $v = P_0 (k_y/\Omega) s\exp(i\theta/\varepsilon)$. Taking $P_0 = 1/|\mathbf{k}|^2$, locally this expression is precisely the spanwise velocity field for a linear monochromatic internal wave with wavenumber $\mathbf{k}$ and steepness $s$. The same is true for the vertical velocity and buoyancy perturbation fields. The streamwise velocity field is instead given by 
% \begin{eqnarray}
%     u = \left[\frac{k_x}{\Omega |\mathbf{k}|^2} +\frac{ik_y}{\Omega^2|\mathbf{k}|^2}\frac{\partial U}{\partial y}\right]s\exp(i\theta/\varepsilon),
% \end{eqnarray}
% where the second term arises due to the term $vU'(y)$ in equation (6) representing wave advection of the streamwise momentum.
The local wave energy density $E=E_K + E_P = |\mathbf{u}|^2/2+|\theta|^2/2$, where $E_K$ and $E_P$ are the kinetic and potential energy densities, is 
\begin{eqnarray}\label{eq:waveenergy}
    E = \frac{s^2}{2}\Bigl[ \overbrace{\frac{1}{k_z^2}}^{O(\epsilon^2)} +\overbrace{\left(\frac{k_y}{|\mathbf{k}|^2\Omega^2}\frac{\partial U}{\partial y}\right)^2}^{O(\epsilon^2 (\partial U/\partial y)^2)} \Bigr]. \label{eq:waveenergy}
\end{eqnarray}
 % In this case, the kinetic and potential energy remain equipartitioned $E_K =E_P$ throughout the motion. However, if the second term in equation (18) becomes large, this is no longer the case. 
It follows from equations \eqref{eq:wave} and \eqref{eq:waveenergy} that wave advection of momentum can result in `excess' kinetic energy contained purely in the streamwise velocity component, here referring to energy not accounted for by the usual wave energy density $s^2/2k_z^2$. The presence of this additional energy has no effect on the dynamics of wave propagation (for example, the group velocity and dispersion relation) at leading order because $Uk_x\sim O(1)$. In particular, this applies to the conservation equation for wave action $\mathcal{A}$ (e.g. \citealp{bretherton1968wavetrains}), which for the case of a wave packet periodic in $x$ and $z$ reads
\begin{eqnarray}\label{eq:waveaction}
    \frac{\partial \mathcal{A}}{\partial t} + \frac{\partial }{\partial y}(c_{g,y}\mathcal{A}) = -\frac{1}{Re}|\mathbf{k}|^2\mathcal{A} \iff \frac{d\mathcal{A}}{dt} = -\mathcal{A}\frac{\partial c_{g,y}}{\partial y} - \frac{1}{Re}|\mathbf{k}|^2\mathcal{A},
\end{eqnarray}
where $\mathcal{A} \equiv s^2/(2 k_z^2)/\Omega$. Here we have also included a viscous term on the right hand side due to \citet{grimshaw1974internal} (see also \citealp{koop1981preliminary, howland2021shear}).

The discussion above suggests that perturbation energy may change due to two distinct mechanisms. Wave advection of momentum leads to a change in the kinetic energy contained entirely in the streamwise velocity component according to the local polarisation relations \eqref{eq:wave}. On the other hand, wave refraction by the horizontal shear modifies the intrinsic frequency and group velocity and hence changes the local wave steepness $s$ via wave action conservation. In the context of the ray-tracing equations, these behaviours may be interpreted as \textit{local} manifestations of the global transient energy growth mechanisms referred to by \citet{bakas2009gravitya,bakas2009gravityb} as growth due to the roll (lift-up) mechanism and to downgradient Reynolds stresses, respectively. Note that these mechanisms operate independently of one another when $Uk_x\sim O(1)$, but may be simultaneously active.

%- \overbrace{2\frac{N^4k_yk_x}{\Omega^3|\mathbf{k}|^4}\frac{\partial U}{\partial y}}^{O(\varepsilon^3)}
% where the last term is $O({\varepsilon^3})$ because $k_x\partial U/\partial y = O(1)$ by assumption. 

% The wave energy density satisfies a 
% that is, the total action within a fixed volume moving at the group velocity $c_{g,y}$ is conserved. 
\subsubsection{Limiting dynamics and instability}
As noted above, when $k_x=0$ it is clear from \eqref{eq:wkbwavenumbers} that the local wave state is unaffected by the shear flow. Thus in this case energy growth occurs purely through wave advection of mean flow momentum. As long as perturbations remain independent of $x$, the total streamwise momentum $U(y)+u(y,z,t)$ behaves simply as a passive scalar in the system. This behaviour is often referred to as the `lift-up' mechanism for shear flow perturbations dating back to the works of \citet{landahl1980note, ellingsen1975stability, moffat1967interaction} and leads to the generation of $y$-vorticity and hence vertical shear via vortex tilting \citep{butler1992three}. Note that $\Omega^2 = k_y^2/|\mathbf{k}|^2$ in this case so that 
\begin{eqnarray}
    E = \frac{s^2}{2}\left[ \frac{1}{k_z^2} +\left(\frac{1}{k_y}\frac{\partial U}{\partial y}\right)^2 \right].\label{eq:waveenergy2}
\end{eqnarray}
Hence the ratio of the bracketed terms in equation \eqref{eq:waveenergy2} is $1+ Fr_h^2 k_z^2/k_y^2 $, where $Fr_h=\partial U/\partial y = (\partial U^*/\partial y^*)/N^*$ is the local horizontal Froude number associated with the background shear flow. Though the analysis here is strictly local, it is worth pointing out that this is an equivalent scaling to that derived for the optimal maximum \textit{total} energy growth for $x$-independent perturbations in an unbounded uniform horizontal shear flow by \citet{bakas2009gravitya}. The greatest perturbation energies result from waves with large horizontal to vertical aspect ratio $k_z/k_y$ in a flow with large $Fr_h$. 
% Note that the transient growth itself is not captured by the ray-tracing theory because it occurs on the fast timescale of phase variations by construction. 
% Additionally, the formation of regions of large vertical shear $\partial u/\partial z$ leads to the possibility of subsequent instability and a breakdown to turbulence. 

More generally, the ratio of the two components of the wave energy density in equation \eqref{eq:waveenergy} forms a dimensionless parameter $F$ defined by
\begin{eqnarray}\label{eq:Dparameter}
    F^2 = \frac{k_y^2k_z^2}{\Omega^4|\mathbf{k}|^4}\left(\frac{\partial U}{\partial y}\right)^2 =\frac{k_z^2}{k_y^2(1+k_x^2/k_y^2)^2}\left(\frac{\partial U}{\partial y}\right)^2 \, ,
\end{eqnarray}
describing the instantaneous proportion of excess kinetic energy due to wave advection of momentum. Note that $F$ varies along rays due to changes in $k_y$ and $\partial U/\partial y$. The evolution may be readily calculated by solving the ray tracing equations \eqref{eq:wkbwavenumbers} numerically. \citet{badulin1985trapping} demonstrate that the wave steepness $s$ increases due to energy extraction from the shear near trapping levels. As suggested by \citet{bakas2009gravitya}, when $k_x<0$ (so that $k_y$ is increasing) and $F$ is small so that wave advection of momentum is relatively unimportant, corresponding amplitude growth may lead to the formation of regions of unstable buoyancy gradients and hence the potential for convective instability. As $F$ increases, the importance of wave advection of momentum increases, which leads to a potentially significant increase in the vertical shear. This can be seen in our analysis by using the local wave polarisation relations \eqref{eq:wave} to compute that $|\partial u/\partial z|^2 +|\partial v/\partial z|^2 = k_z^2/|\mathbf{k}|^2 +F^2$.

Dynamically, the tendency for vertical shear to destabilise the flow is characterised by a local gradient Richardson number 
\begin{eqnarray}
    Ri_g(\mathbf{x},t) = \frac{1 + \partial \theta/\partial z}{S^2} = \frac{1 + \partial \theta/\partial z}{(\partial u_{\beta}/\partial z)^2},
\end{eqnarray}
defined here based on a locally parallel shear flow $u_{\beta}= u\sin\beta +v\cos\beta$ obtained from the projection of the horizontal velocity field onto a chosen azimuthal direction $\beta$ in which unstable perturbations are assumed to grow. A criterion of $Ri_g<1/4$ somewhere in the flow is often adopted as an indicator of the potential for shear instability based on the classical Miles-Howard theorem \citep{miles1961stability,howard1961note}, though the formal assumptions required for the theorem to be applicable are invalid even for the simple linear wave state considered here. Nonetheless, $Ri_g<1/4$ has proven useful as a qualitative indicator of flow instability in a wide range of flows. For a linear internal gravity wave given by the polarisation relations \eqref{eq:polarisation}, it may be shown (e.g. \citealp{thorpe1999breaking}) that, for statically stable configurations $s<1$, the minimum value of $Ri_g$ is $1/2$, and so shear instability is not expected. However, when the wave interacts with a horizontal shear flow such that the streamwise velocity field is modified according to equations \eqref{eq:wave}, significant changes to $Ri_g$ may be expected.

In the limit $k_x=0$, the horizontal velocity components $u$ and $v$ differ by a phase of $\pi/2$. In this case, an analogy with inertia-gravity waves in rotating stratified flows provides an interesting point of comparison.  In particular, the polarisation between these two components is the same as a linear inertia-gravity wave propagating in the $y$-$z$ plane under the association $\partial U/\partial y \leftrightarrow -f$ (importantly however, the same is not true of the dispersion relation). Unlike those in systems with $f=0$, such waves may exhibit $Ri_g<1/4$ for $s<1$, with the resulting dynamics closely resembling classical parallel shear instability \citep{lelong1998inertiaa}. The differences between the resulting energetics and those associated with convective instabilities may be significant \citep{thorpe2018models}. For $F\neq 0$, the local gradient Richardson number for the horizontally sheared system considered here is calculated to be
\begin{eqnarray}\label{eq:localri}
    Ri_g(\mathbf{x},t) = \frac{1+s\cos\phi}{F^2s^2 [\cos\phi \sin\beta +(k_x/k_y)(\sin\phi\sin\beta)/R +(\sin\phi\cos\beta)/R]^2},
\end{eqnarray}
where $\phi = \mathbf{k}\cdot \mathbf{x} - \Omega t$ is the local wave phase and $R= (\partial U/\partial y)/\Omega$. It is demonstrated in Appendix \ref{sec:appa} that the minimum value of $Ri_g$ is very close to (and in some cases exactly equal to) $(1-s)/F^2s^2$ when this quantity is smaller than $0.5$. This value is achieved at $\beta=\pi/2$ and $\phi = \pi$, corresponding to perturbations transverse to the direction of wave propagation (c.f. \citealp{lelong1998inertiaa}). It is therefore apparent that $F$ may be useful as a proxy for shear instability. Indeed, because $k_z/k_y$ represents a horizontal to vertical wave aspect ratio, the expression \eqref{eq:Dparameter} takes the form of an inverse Richardson number. 

% as for the plane inertia gravity wave studied in \citet{lelong1998inertiaa},  

The physical meaning of $F$ breaks down when $k_y$ becomes small such that the WKBJ approximation is no longer valid. However, for rapid growth in the streamwise velocity field (i.e. large $\partial U/\partial y$), the small-amplitude linear approximation may break down before this point. In the special case $k_x=0$, Lagrangian conservation of total streamwise momentum limits the maximum amplitude of the streamwise velocity to be at most $\Delta u$ and the expression for $F$ is certainly meaningless beyond this limit. With these restrictions in mind, we consider the ray tracing equations under some representative cases to illustrate the variability in dynamics that can occur due to the influence of horizontal shear. We fix $k_z=1$ and choose values of $k_x$, $k_y$, $\Delta u$ and $h$ to produce scenarios where the wave is trapped ($k_x<0$), reflected ($k_x>0$) and unaffected by the shear ($k_x=0$), consistent with the direct numerical simulations described in section~\ref{sec:simulations}. The ray tracing equations \eqref{eq:wkbraypaths} and \eqref{eq:wkbwavenumbers} are solved numerically alongside equation \eqref{eq:waveaction} governing the evolution of wave action $\mathcal{A}$. Local values of the wave steepness $s$ are computed from $s^2 = 2\mathcal{A}\Omega k_z^2 $. 

\begin{figure}
\includegraphics[width=\textwidth]{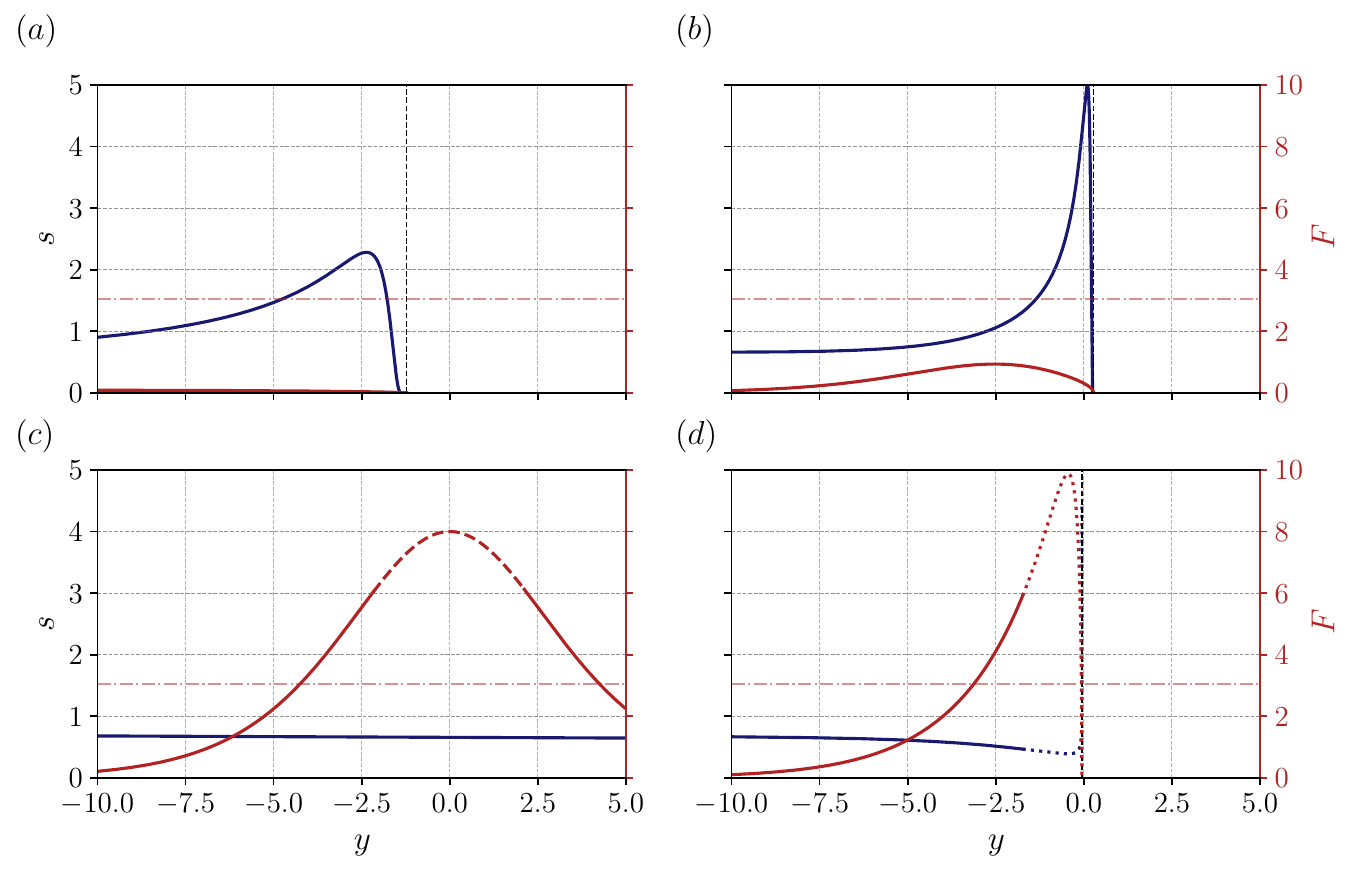}
\caption{Plots of the wave steepness $s$ derived from the local wave action $\mathcal{A}$ (dark blue, left axes) and the parameter $F$ defined in equation \eqref{eq:Dparameter} (dark red, right axes) along rays in the $y$ direction for parameter values corresponding to selected simulations from table \ref{tab:simulation-results}. $(a)$ F0.1s2.28(T) and $(b)$ F1.0s5.0(T) (wave trapping), $(c)$ F6.0s0.75\_kx0 (no refraction) and $(d)$ F6.0s0.75 (wave reflection). Lines become dashed in panel $c)$ where the perturbation velocity exceeds $\Delta u$ and dotted in panel $(d)$ where $k_y<0.1$. Vertical dashed lines indicate a trapping level in $(a), (b)$ and a turning level in $(d)$. Horizontal dot-dashed lines represent the value of $F$ corresponding to a minimum $Ri_g$ smaller than $1/4$ (assuming $s=0.75)$.   \label{fig:wkb} }
\end{figure}

Values of $s$ and $F$ along rays are shown in figure \ref{fig:wkb} for four representative cases. The first two cases have $k_x<0$ and both exhibit wave trapping, but differ in the value of $\Delta u/h$ by approximately an order of magnitude. The third case is the special case $k_x=0$ (no refraction) and the fourth case has $k_x>0$ such that wave reflection is predicted. It is important to keep in mind that the WKBJ theory is formally invalid near trapping and turning levels.

We first consider the behaviour of $s$. When $k_x<0$, $k_y$ increases monotonically along rays, leading to a decrease in the group velocity as the trapping level at $y\approx 0$ is approached. This leads to the significant increase in the wave steepness observed in panels $(a)$ and $(b)$ before viscosity acts to damp further energy growth, consistent with the results of \citet{badulin1985trapping}. When $k_x=0$, no refraction takes place and wave energy is conserved along rays in the inviscid limit. This corresponds to the approximately constant value of $s$ seen in panel $(c)$, which decreases only due to viscosity. When $k_x=0.05$ in panel $(d)$ such that a turning level exists at $y\approx 0$, $s$ similarly remains approximately constant as the turning level is approached, until the linear WKBJ approximation becomes invalid at $y\approx -2$. The parameter $F$ exhibits distinctly different behaviour to that of $s$. When $k_x<0$, the growth in $F$ depends on $\partial U/\partial y$. When $\Delta u/h = 0.125$ is small, as in panel $(a)$, the dynamics are dominated by wave refraction and $F$ remains close to $0$. For larger $\Delta u/h=1$ (panel $b$), $F$ is able to grow to $O(1)$ before growth in $k_y$ eventually dominates. When $k_x=0$ (panel $c$), energy growth is due entirely to the lift-up mechanism, demonstrated by the increase in $F$ as the wave approaches the center of the shear layer. A similar behaviour is observed for $k_x>0$ (panel $d$). Note that, for a given $k_{x0}$, $k_{z0}$ and $\Delta u/h$, the maximum value of $F$ is obtained when $k_y^2=k_{x0}^2<k_{y0}^2$, explaining why a larger maximum is predicted for the reflected case where $k_y$ may vary. 
\subsubsection{Summary}
Before continuing, it is worth briefly summarising the use of the ray-tracing theory described above for predicting properties of the local wave state prior to breaking. For a monochromatic wave packet with wavevector $\mathbf{k}_0$ and steepness $s_0$ initially centered at $y=y_0$ in the horizontal shear flow $U(y)$ given by \eqref{eq:backgroundflow}, the ray-tracing equations \eqref{eq:wkbraypaths} and \eqref{eq:wkbwavenumbers} are solved numerically alongside equation \eqref{eq:waveaction} for the wave action $\mathcal{A}$. The resulting local values of $k_y$ and $\mathcal{A}$ along rays are then used to compute corresponding values of $s$ and $F$ describing the relative importance of buoyancy overturns versus enhanced vertical shear in driving instability, as inferred from the local wave state defined by the polarisation relations \eqref{eq:wave}.

When $k_x<0$ and the wave is trapped or refracted strongly by the horizontal shear flow such that $s$ becomes larger than 1, if $F\ll 1$ remains small then mixed shear/convective instability associated with a statically unstable local buoyancy profile is expected. If $F\gtrsim 1$, wave advection of momentum leads to increased vertical shear that may modify the properties of instability. On the other hand, when $k_x\geq 0 $, trapping does not occur (so that the wave passes through the shear layer or is reflected) and $s$ remains smaller than $1$. In such cases, $F\gtrsim 1$ may lead to growth of local horizontal velocity perturbations that are sufficient to destabilize the flow to vertical shear instability alone. %We reiterate that the analysis above is at best  heuristic, being strictly valid only for small amplitude wave perturbations in the WKB approximation. However, we will see that it provides a useful framework for interpreting the fully nonlinear direct numerical simulations described below. 

\section{Numerical simulations}\label{sec:simulations}

The linear theory in the previous section suggests that a range of wave-breaking behaviours are possible in a horizontal shear layer, depending on the initial flow parameters. However, this analysis is at best heuristic, being strictly valid only for small-amplitude wave perturbations in the WKBJ approximation. To further understand the flow evolution, we perform a series of complementary fully nonlinear direct numerical simulations (DNS) intended to span the range of predicted breaking behaviours. Simulations are performed using the multi-parallel pseudo-spectral channel flow solver \textsc{Diablo} \citep{Taylor08}, which employs a second-order finite difference discretization in the wall-bounded direction, here taken to be in the direction of horizontal shear. The code uses a third-order mixed implicit/explicit Runge-Kutta/Crank-Nicolson scheme for time marching, with a 2/3 dealiasing rule applied to the nonlinear terms. 

\subsection{Overview}

Under the expectation that $k_z$ remains constant throughout the wave-shear interaction, it is convenient to take $K^* = k^*_{z0}$ in the non-dimensionalisation outlined in \S\ref{sec:ivp}. In this way, the dimensionless initial wavenumber vector $\mathbf{k}_0 = (k_{x0}, k_{y0}, 1)$ has corresponding frequency $\Omega_0$ satisfying $\Omega_0^2 = (k_{x0}^2+k_{y0}^2)/(1+k_{x0}^2+ k_{y0}^2)$, and the initial (maximum) wave energy density $E_0 = s_0^2/(2k_{z0}^2)=s_0^2/2$. Most of the simulations presented here have $k_{y0}=0.25$, with values of $k_{x0}$ then selected to span a range of predicted trapping and reflecting scenarios. The dimensionless initial frequency $0.125<\Omega_0<0.38$ (corresponding to propagation angles of $13^{\circ}$-$21^{\circ}$ between $\mathbf{k}_0$ and the horizontal) is large enough such that the neglect of planetary rotation is reasonable (under the assumption $f\ll N$) without being so large that waves quickly become unstable to a parametric instability \citep{lombard1996breakdown}. We note these values are within the estimated range of high frequency waves observed in the ocean thermocline by \citet{alford2000observations}. The packet width scale is chosen as $a=6\pi$, with the envelope centered at $y_0 = -2a$. Simulations are initialised with $\mathbf{u}(\mathbf{x},t=0)=\mathbf{U}(y)+\mathbf{u}_{\mathrm{wave}}(\mathbf{x})$ and $b=b_{\mathrm{wave}}(\mathbf{x})$ from \eqref{eq:wavepacket} and \eqref{eq:backgroundflow}. Additionally, white noise of amplitude $10^{-5}$ is added to each component of the velocity field in order to trigger instabilities leading to turbulence.

\subsection{Inclined computational domain}\label{sec:incline}
%not totally happy with this section right now
In order to allow for waves with $k_{x0}\ll 1$, it is convenient to perform simulations in an inclined reference frame $(x,y,z)\to (\hat{x}, y, \hat{z}) \equiv (x\cos\alpha +z\sin\alpha, y, z\cos\alpha - x\sin\alpha)$ obtained by rotation in the $x$-$z$ plane by an angle $\alpha = -\arctan(k_{x0}/k_{z0})$. In this frame, $k_{x0}x+k_{z0}z = \tilde{k}_{z0}\tilde{z}$, where $\tilde{k}_{z0} = k_z/\cos\alpha$. 
%In order to allow for waves with $k_{x0}\ll 1$, it is convenient to perform simulations in an inclined reference frame $(x,y,z)\to (\tilde{x}, \tilde{y}, \tilde{z})$ obtained by rotation in the $x$-$z$ plane by an angle $\alpha = -\arctan(k_{x0}/k_{z0})$. It follows that $(\tilde{x}, \tilde{y}, \tilde{z}) = (x\cos\alpha +z\sin\alpha, y, z\cos\alpha - x\sin\alpha)$, and hence that $k_{x0}x+k_{z0}z = \tilde{k}_{z0}\tilde{z}$ where $\tilde{k}_{z0} = k_z/\cos\alpha$. 
As a result, the initial wave fields $[\hat{u}, \hat{v}, \hat{w}, \hat{b}]$ are proportional to $\exp(\mathrm{i}k_{y0} y + \mathrm{i}\tilde{k}_{z0} \hat{z})$ and the initial condition is independent of $\hat{x}$. This removes the restriction that the domain length in the $x$-direction must be an integer multiple of $2\pi/k_x$ and allows the possibility for arbitrarily small $k_{x0}$ to be investigated. Further details may be found in appendix \ref{sec:appb}. A similar approach for studying the instability of plane periodic internal waves is adopted by \citet{fritts2013gravityb,fritts2013gravitya, fritts2003layering}. As explained by these authors, it is important to note that due to the numerical discretisation and finite domain size, the permitted modes of instability are not the same between the original and inclined coordinate frames. In practice however, provided the largest-scale modes are smaller than the domain size and the resolution is sufficiently high, the subsequent dynamics are expected to be similar between the two. Indeed, in general we find that the most energetic modes of instability in the $\hat{x}$ direction have wavenumber $k_x \gtrsim 1$. Additionally, some preliminary simulations with $k_{x0}=-1/3$ (not shown) were performed in physical coordinates and compared to calculations performed in the inclined frame, with instability and turbulent breakdown being qualitatively and quantitatively very similar between the two. 

The domain length in the periodic $\hat{x}$ and $\hat{z}$ directions is taken to be $L_x=L_z=8\pi$, whilst a larger extent $y\in (-100,100)$ is used in the direction of shear to accommodate the initially separated wave packet and shear layer.  The grid is evenly spaced in the $\hat{x}$ and $\hat{z}$ directions with $N_x=N_z=648$ grid points in physical space. Polynomial grid stretching is used in the finite difference $y$ direction so that the finest resolution is achieved at the location $y=0$ of maximum shear in the centre of the domain and is equivalent to the grid spacing in the $\hat{x}$ and $\hat{z}$ directions. A total of $N_y=1177$ grid points are used in the $y$ direction, with $577$ of these points concentrated in the central region $-12.5 \leq y \leq 12.5$. The grid resolution is selected such that grid scale remains at most 2.5 times larger than the Kolmogorov length scale $L_K = \max (\overline{\varepsilon} Re_w^3)^{-1/4}$ throughout each simulation, where $\overline{\varepsilon}$ is the kinetic energy dissipation rate averaged over $\hat{x}$ and $\hat{z}$ (cf. \citealp{kaminski2019stratified}). Free-slip/no-flux boundary conditions are applied at the walls $y=\pm 100 $ ($v=0$, $\partial \phi/\partial y=0$ for $\phi=u,w,b,p$). A sponge layer is employed in regions of width $5$ dimensionless units of length adjacent to the boundaries, within which perturbations are damped to zero quadratically in order to prevent wave reflection back into the centre of the domain.

Importantly, though the simulations presented here are performed in the inclined reference frame, output fields shown in what follows have been transformed back into the physical coordinate system. 

\subsection{Parameter space}

\begin{table}
\centering
\begin{tabular}{ccccccccc}
\textbf{Simulation} & \boldmath{$\Delta u$} & \boldmath{$h$} & \boldmath{$k_{x0}$} & \boldmath{$s_0$} & \boldmath{$Fr_h$} & \boldmath{$Re_w$} & \boldmath{$F_{\mathrm{max}}$} & \boldmath{$s_{\mathrm{max}}$}\\
\hline
F0.1s0.76(T) & 2 & 16 & -0.33 & 0.25 & 0.125 & 600 & 0.1 & 0.76 \\

F0.1s2.28(T) & 2 & 16 & -0.33 & 0.75 & 0.125 & 600 & 0.1 & 2.28 \\

% U2H4(T) & 2 & 4 & -0.33 & 0.75 & 0.5 & 600 & 0.121 \\

F1.0s1.66(T) & 4 & 4 & -0.17 & 0.25 & 1.0 & 300 & 1.0 & 1.66 \\

F1.0s5.0(T) & 4  & 4 & -0.17 & 0.75 & 1.0 & 300 & 1.0 & 5.0\\

% U8H4(T) & 8 & 4 & -0.08 & 0.75 & 2.0 & 150 & 4.343 \\

F3.0s0.25 & 2 & 4 & 0.08 & 0.25 & 0.5 & 600 & 3.0 & 0.25\\

F3.0s0.75 & 2 & 4 & 0.08 & 0.75 & 0.5 & 600 & 3.0 &  0.75\\

F6.0s0.25 & 4 & 4 & 0.05 & 0.25 & 1.0 & 300 & 6.0 & 0.25 \\
F6.0s0.75 & 4 & 4 & 0.05 & 0.75 & 1.0 & 300 & 6.0 & 0.75 \\

F6.0s0.75\_kx0 & 4 & 4 & 0 & 0.75 & 1.0 & 300 & 6.0 & 0.75 \\
\end{tabular}
\caption{Table of simulations with parameters (as defined in the text) listed, where the naming convention specifies $F_{\mathrm{max}}$ and $s_{\mathrm{max}}$, with (T) referring to simulations where the incident wave is trapped. All simulations have $a=6\pi$, $y_0=-2a$ and $Pr=1$. Note all simulations have $k_{y0}=0.25$ except for simulation F6.0s0.75\_kx0, which has $k_{y0}=0.125$. The shear-layer Froude number $Fr_h$, wave Reynolds number $Re_w$, maximum kinetic to potential energy ratio $F_{\mathrm{max}}$ and maximum wave steepness $s_{\mathrm{max}}$ are defined in the text.}\label{tab:simulation-results}
\end{table}
Values of $k_{x0}$, $\Delta u$, $h$ and $s_0$ are selected to span a range of wave-shear interaction dynamics. When negative, the value of $k_{x0}$ is chosen such that a trapping level exists at $y=y_t=0$ according to equation \eqref{eq:yt_implicit} above. Analogously, when positive, $k_{x0}$ is chosen such that a turning level exists at the same location. The horizontal Froude number $Fr_h = \Delta u/h $ of the initial shear flow \eqref{eq:backgroundflow} varies from $0.125$ to $1$. Because the wave extracts kinetic energy from the background shear prior to breaking, a Reynolds number $Re_I$ more appropriate for describing the resulting turbulence may be defined using the shear flow velocity scale $\Delta u$ rather than the wave velocity scale. As a result, 
\begin{eqnarray}
    Re_I \equiv \frac{\Delta u^*}{k_{z0}^* \nu^*} = Re_w\Delta u 
\end{eqnarray}
is chosen as $Re_I=1200$ for all simulations. Most simulations use an initial wave steepness of $s_0=0.75$, though, similar to \citet{winters1994three} and \citet{howland2021shear}, we also perform some simulations at a smaller amplitude of $s_0=0.25$ to investigate the effect of wave amplitude on the resulting instability. Additional simulations with $\Delta u =0 $ (not shown) were performed to verify that the propagating initial wave packet remains stable in the absence of the shear flow. Thus the dynamics of wave instability studied below are instigated by wave-shear interaction and not due solely to parametric self-interaction instability \citep{lombard1996instability} or the interaction of the wave with its amplitude envelope \citep{sutherland2006weakly}.

A list of DNS is displayed in table \ref{tab:simulation-results}. Included are the maximum values of $F_{\mathrm{max}}$ and $s_{\mathrm{max}}$ of the parameter $F$ defined in equation \eqref{eq:Dparameter} and local steepness $s$ computed along rays using the ray tracing equations outlined above. Larger values of $F_{\mathrm{max}}$ correspond to an increase in the importance of wave advection of streamwise momentum resulting in excess perturbation kinetic energy, whilst values of $s_{\mathrm{max}}>1$ indicate the potential for convective instability. We stress that the precise magnitudes of these parameters are valid only for linear wave perturbations that satisfy the assumptions required for validity of the WKBJ approximation. However, it will be shown that $F_{\mathrm{max}}$ remains useful as a predictor for the partition between kinetic and potential energy in the system even for the strongly nonlinear waves considered here. For the purposes of comparison, we note that in our non-dimensionalisation, simulations 1-4 of \citet{staquet2002transport} have $\Delta u = 4$, $h=4$, $k_x=-0.25$, $k_y=0.25$, $a=4$, $y_0=-8$, $s_0=0.3$ and $Re_w\approx 2500$.

%In the ocean, it is expected that horizontal wavelengths are much larger than vertical wavelengths so that, in our non-dimensionalisation, we should ideally have $k_x,k_y\ll 1$. Wave packets with very small $k_y$ must be sufficiently wide for the expected group velocity to be physically meaningful and therefore require a prohibitively large domain size if the full range of scales of wave and turbulent motions are to be resolved. As a compromise, a moderately small value of 

% In the results that follow, visualisations of flow fields from the simulations are shown in the inclined frame of reference. However, the flow fields themselves are transformed back into the standard frame of reference with gravity in the vertical direction (e.g. $u(\tilde{x},\tilde{y},\tilde{z})$). 
%Maybe add details about the rotation in an appendix? 

%The linear framework above is now used to construct a parameter space for fully nonlinear direct numerical simulations that are expected to span a range of breaking behaviours. 

\section{Results}\label{sec:results}
\begin{figure}
\includegraphics[width=\textwidth]{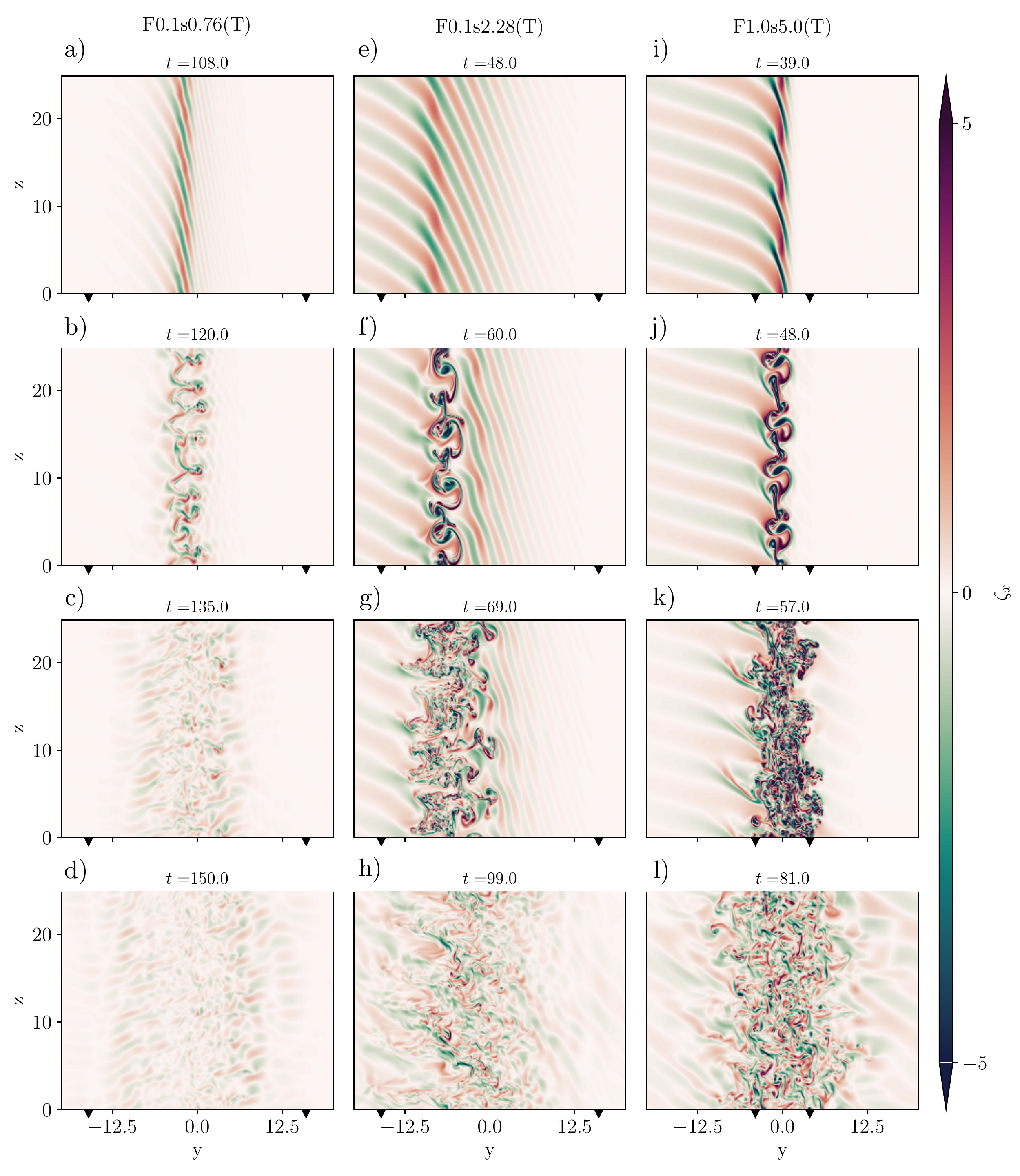}% Here is how to import EPS art
\caption{\label{fig:trapped_panels} Snapshots in a vertical plane of the central region $-15<y<15$ taken at successive time points during the flow evolution for simulations $(a)$-$(d)$ F0.1s0.76(T); $(e)$-$(h)$ F0.1s2.28(T); $(i)$-$(l)$ F1.0s5.0(T). Colours illustrate the streamwise vorticity field $\zeta_x$. Triangle markers along the horizontal $y$-axis are located at $y = \pm \Delta u$, indicating the approximate width of the shear layer.  }
\end{figure}
We now present results obtained from the DNS outlined in the previous section, starting with a qualitative picture of the wave-shear interaction. For ease of communication, the results are split into simulations with $k_x<0$ in which the wave is trapped, and those with $k_x\geq 0$, where the wave either reflects or passes through the shear layer. As detailed above, the dynamics are expected to be characteristically different between these cases. 

\subsection{Evolution for $k_x<0$ (wave trapping)}
\begin{figure}
\includegraphics[width=\textwidth]{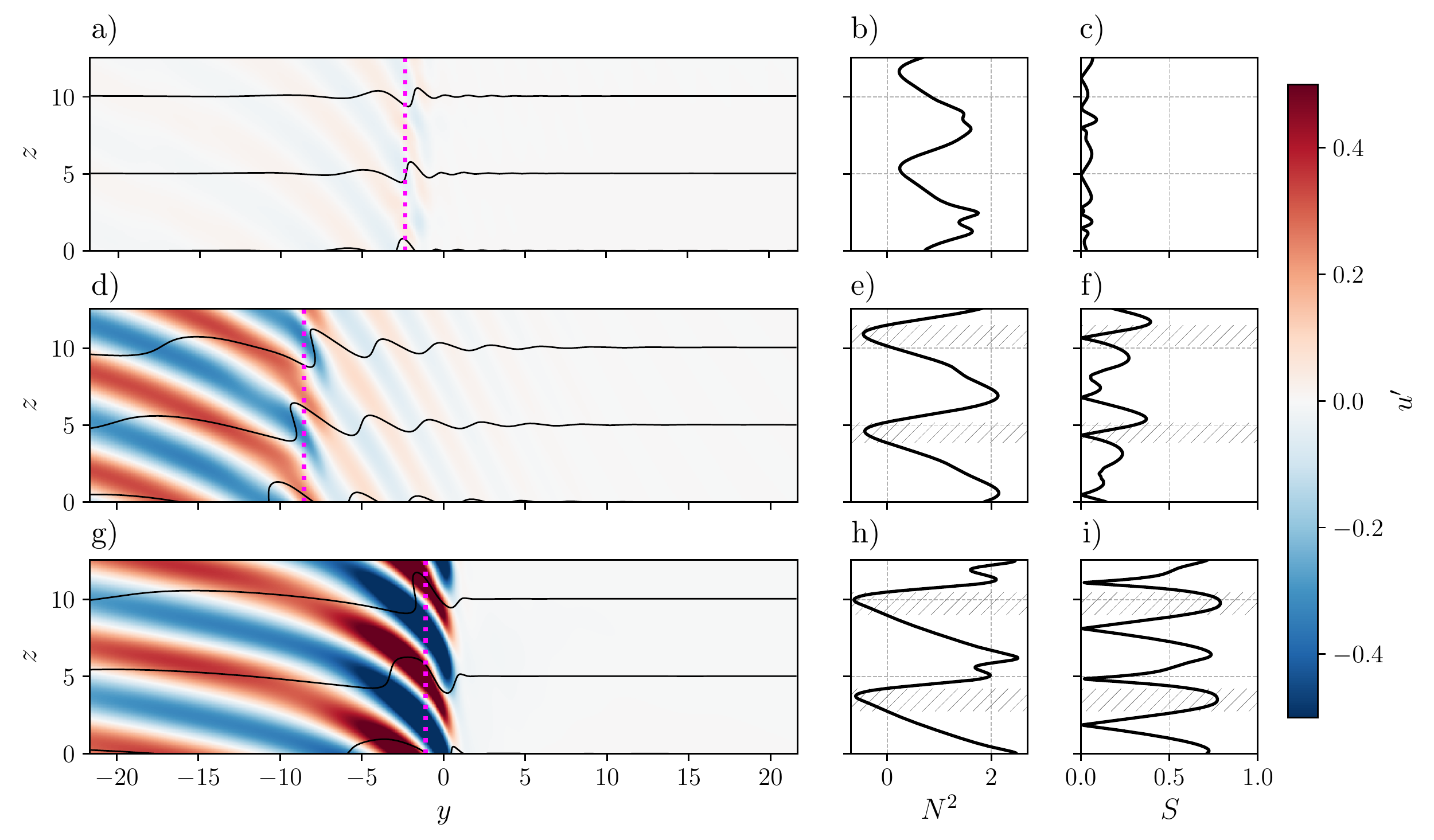}% 
\caption{\label{fig:4} $(a)$, $(d)$ and $(g)$ show vertical slices of the perturbation streamwise velocity field for simulations F0.1s0.76(T), F0.1s2.28(T) and F1.0s5.0(T) taken at times corresponding to the panels in the top row of figure \ref{fig:trapped_panels}. Solid black lines are evenly spaced isopycnals (surfaces of constant buoyancy $b=z+\theta$). Vertical profiles of the local buoyancy frequency $N^2 = 1 +\partial \theta/\partial z$ and vertical shear $S$ are shown in the panels to the right. The locations at which these profiles are taken are indicated by the dashed magenta lines. Hatched regions show vertical locations where $N^2<0$.}
\end{figure}
The evolution of the wave packet in the $(y,z)$-plane is shown in figure \ref{fig:trapped_panels} by plotting contours of the streamwise vorticity field $\zeta_x = \partial w/\partial y - \partial v/\partial z$.  Results from three representative simulations with $k_x<0$ are presented to illustrate similarities and differences in the dynamics between cases with varying $s_{\mathrm{max}}$ and $F_{\mathrm{max}}$. Snapshots are taken at times showing the steepening of the incident wave due to refraction in the shear layer, the emergence of instabilities, the subsequent development of small-scale turbulent structures with enhanced vorticity and the eventual decay of these motions. Note that simulations shown in the second and third columns have the same initial parameter values as the results from the linear calculations shown in figure \ref{fig:wkb}$(a)$ and $(b)$. 

% Indeed, the refraction of the wave as it enters the shear layer is apparent in all cases, with contours of $\zeta_x$ steepening towards the center of the domain. However, clear departures from the expected approach to the trapping level are visible even for the smaller amplitude wave in F0.1s0.76(T). 

When $s_0=0.25$ (first column), distortion of the refracted wave packet is visible in panel $(a)$ at time $t=108$ (approximately 17 dimensionless buoyancy periods). Shortly afterwards, in panel $(b)$, small-scale vorticity structures emerge and propagate rapidly across the trapping level. Disordered small scale motions develop in the center of the shear layer but are quickly damped by viscosity by time $t=150$ in panel $(d)$ at the chosen Reynolds number owing to their relatively small amplitude. Nonetheless, small-scale internal waves, first seen in panel $(c)$, are generated at the edge of the central turbulent region and radiate outward in both directions in a manner qualitatively similar to the results of \citet{maffioli2014evolution}. For a larger initial wave amplitude $s_0=0.75$ but keeping all other parameters the same (second column), instability develops much earlier, and at a considerable distance from the predicted trapping level. Without appealing to the linear theory, there is no obvious indication as to the location of the trapping level from the vorticity snapshots, with a small amount of wave energy clearly able to penetrate through $y=0$. Despite these differences, the subsequent evolution of the instability itself is somewhat similar between the two cases, with the formation of localised vorticity structures that propagate towards and across the center of the shear layer and the generation of internal waves by turbulence. Likewise, the picture of instability remains similar when the shear layer width is decreased by a factor of four and the magnitude of the velocity is increased by a factor of two (third column), though turbulence is clearly most energetic in this case. The qualitative behaviour in this case is also consistent with results obtained by \citet{staquet2002transport} for comparable initial values of $\Delta u$, $h$ and $k_{x0}$.

% In the context of the linear analysis from section 2 above, we now analyse in detail the properties of instability prior to the development of the coherent vortex structures and subsequent breakdown to turbulence.
\begin{figure}
\includegraphics[width=\textwidth]{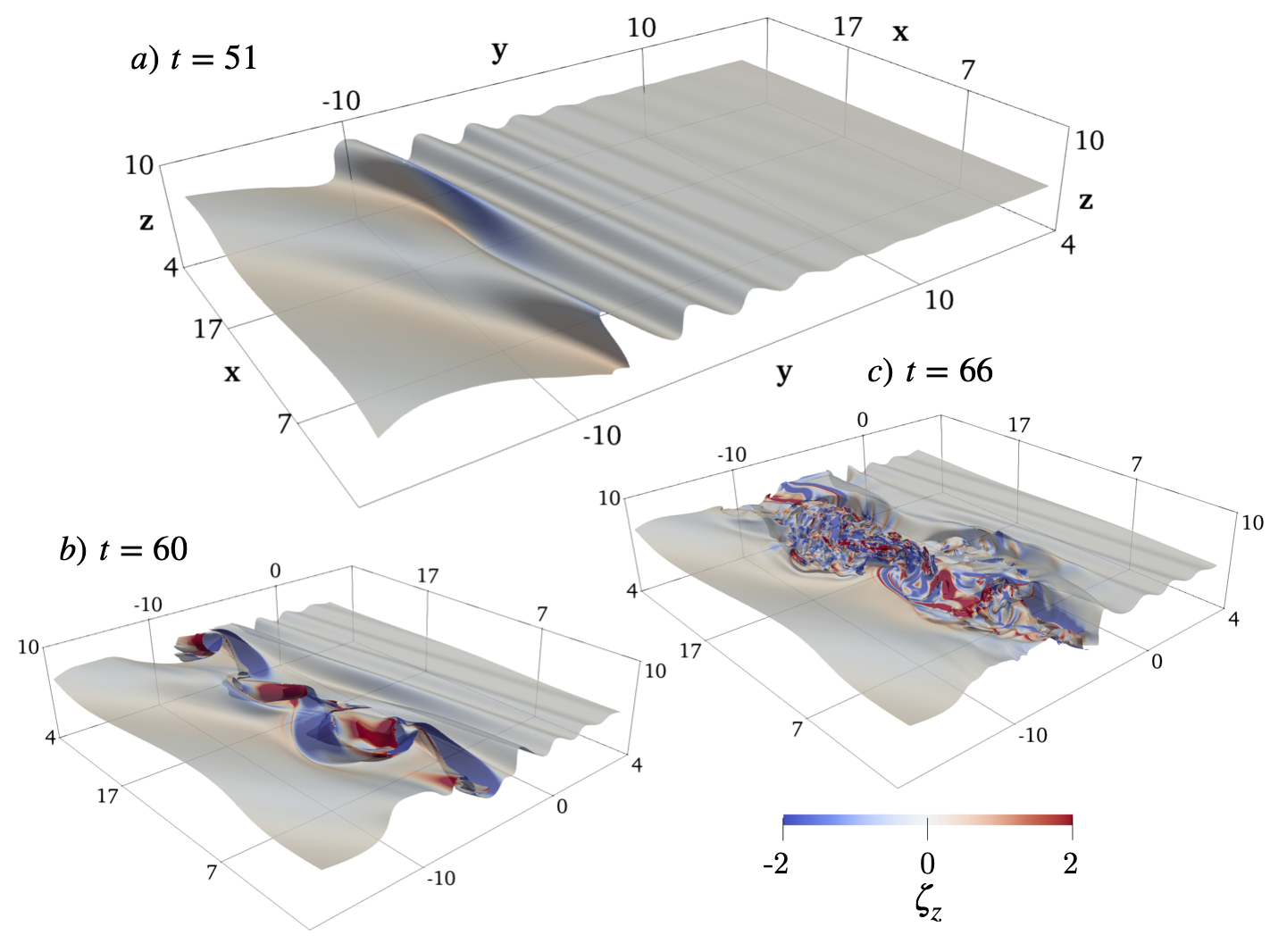}% Here is how to import EPS art
\caption{\label{fig:5} 3D renderings of the $b = 6.5$ isopycnal from simulation F0.1s2.28(T) at successive times during the flow evolution. Colours indicate the magnitude of the vertical vorticity field $\zeta_z$.}
\end{figure}
Snapshots showing the local wave state prior to breaking are shown in figure \ref{fig:4}. Colours show the perturbation streamwise velocity field $ u' = u - \overline{u{}} $, where an overline represents an average in $x$ and $z$. Isopycnals are included to highlight overturning regions. Local vertical profiles of the buoyancy gradient $N^2(z)$ and vertical shear (of the streamwise velocity) $S(z) = |\partial u/\partial z|$ taken at the $y$ location of the maximum velocity amplitude are plotted in the right-hand panels. The results are qualitatively consistent with the predictions from figure \ref{fig:wkb}. In particular, the refraction of the wave by the shear leads to an increase in steepness $s$, as seen from the isopycnals. When the initial amplitude is sufficiently large, local regions of overturning develop (hatched regions in the profiles). It is worth noting the linear theory predicts values of $s>1$ only for parameter values corresponding to the bottom two rows, which is seen to be consistent with the DNS. Larger values of $s$ are naturally associated with increased vertical shear. More important for the shear however is the advection of momentum when $\partial U/\partial y$ is large, indicated by the large values of $u'$ and corresponding enhanced vertical shear seen in the bottom row of figure \ref{fig:4}. This enhanced vertical shear may be associated with the larger value of $F_{\mathrm{max}}$ in simulation F1.0s5.0(T). Notably however, $Ri_g<1/4$ only when $N^2<0$: there are no regions of the flow susceptible to pure shear instability. This is also consistent with figure~\ref{fig:wkb}(b), which shows that although $F$ increases, it never becomes large enough to support a corresponding minimum $Ri_g<1/4$ whilst $s<1$.

%The analysis above suggests the behaviour of instability for the large wave amplitude scenarios shown in figure \ref{fig:4} is controlled by the convective component. 
The three-dimensional behaviour of the wave breaking instability is shown in figure \ref{fig:5} for simulation F0.1s2.28(T). The development of convectively unstable overturns is clear in panel $a)$. At low frequencies, convectively unstable waves break via the production of vorticity that emerges in the form of roll-like structures in the direction parallel to wave crests \citep{howland2021shear}. This behaviour has been explained by analysing perturbations to an assumed quasi-steady background state representing the wave at the point of overturning \citep{winters1992instability,parker2021optimal}. In general, perturbation energy is extracted from the wave by a mixture of vertical shear production and convection via the buoyancy flux. In this study however, it is less clear that the locally large frequency $\Omega\sim N$ expected in the case of wave trapping satisfies the quasi-steady assumption. Indeed, the vortex structures seen in panels $(e)$ and $(i)$ of figure \ref{fig:trapped_panels} and in panel $(b)$ of figure \ref{fig:5} appear unique to the horizontal shear scenario.
% Indeed, the breaking behaviour seen in figure \ref{fig:5}$b)$ is characteristically different to the case of low frequency waves. 
Nonetheless, though the primary instability is fundamentally different to the classical quasi-steady convective wave instability, it is expected that the secondary instability of the streamwise vortices exhibits mixed shear/convection properties \citep{caulfield2000anatomy}. The resulting three-dimensionalisation results in the rapid collapse of the convectively unstable region and development of small-scale turbulent structures seen in panel $(c)$. 
\begin{figure}
\includegraphics[width=\textwidth]{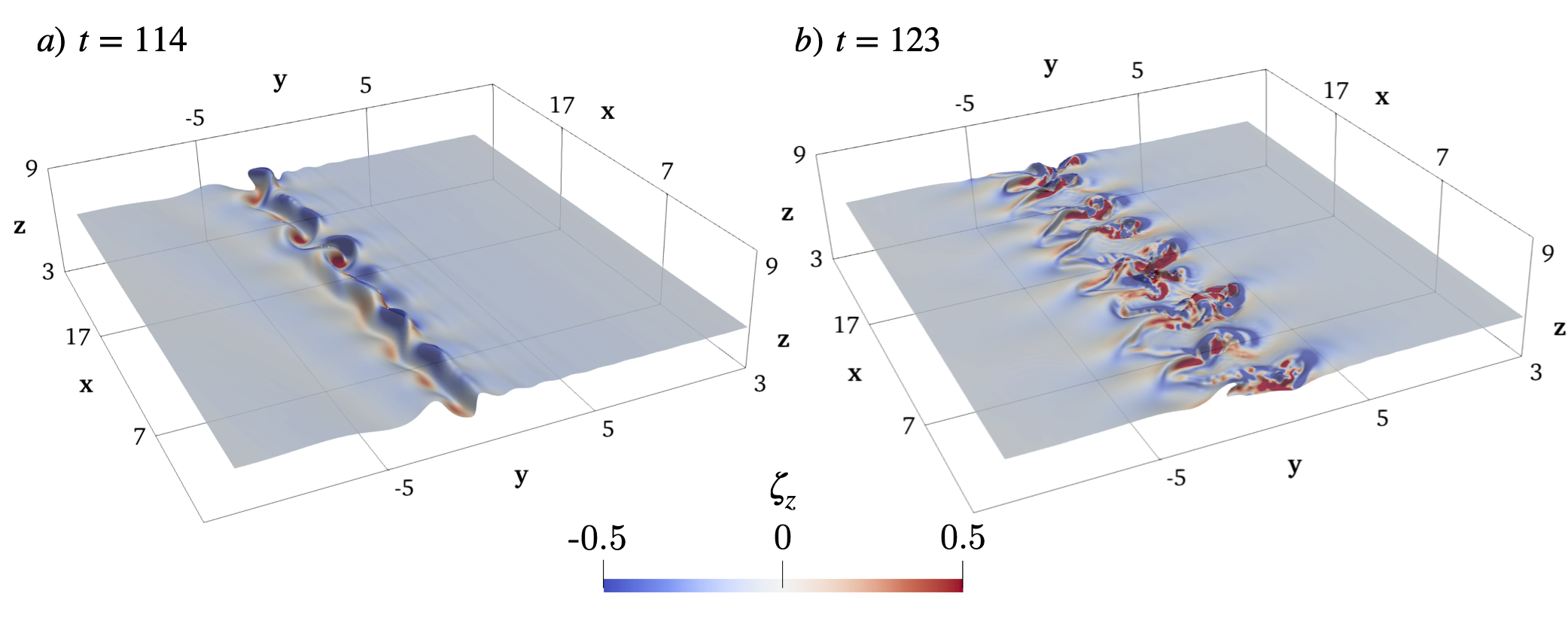}% Here is how to import EPS art
\caption{\label{fig:6} The same three-dimensional rendering as in figure \ref{fig:5} but for simulation F0.1s0.76(T), at two successive time points during the flow evolution.}
\end{figure}

%Consistent with this picture, the regular streamwise defects in the vortex seen in panel $(b)$ are similar in structure to secondary instabilities that emerge in both Kelvin-Helmholtz billows and propagating Holmboe vortices during turbulent breakdown (see, e.g. \citealp{olsthoorn2023dynamics}). 

It is tempting to conclude that turbulent breakdown in the case of wave trapping is ultimately due to a primarily convective instability. However, the case of a smaller-amplitude wave is not consistent with this picture. Figure \ref{fig:6} shows a 3D visualisation of the breaking process for simulation F0.1s0.76(T) with a smaller initial wave amplitude, where it is seen in panel $(a)$ that three-dimensional structure develops \textit{prior} to the overturning of isopycnals. It will be shown in \S\ref{sec:momentum} that wave trapping results in the formation of a narrow velocity jet in the streamwise $x$-direction that is colocated with the region of large isopycnal displacements. Perturbations may then extract energy from the horizontal shear associated with the jet, potentially leading to the generation of vertical vorticity seen in panel $(a)$ and the subsequent breakdown in panel $(b)$. It is worth noting that this behaviour is similar in character to the instability mechanism proposed by \citet{wang2020nonlinearii} for forced critical layers in a stratified uniform horizontal shear flow.
\subsection{Evolution for $k_x\geq 0$}\label{sec:waveadvection}
 \begin{figure}
\includegraphics[width=\textwidth]{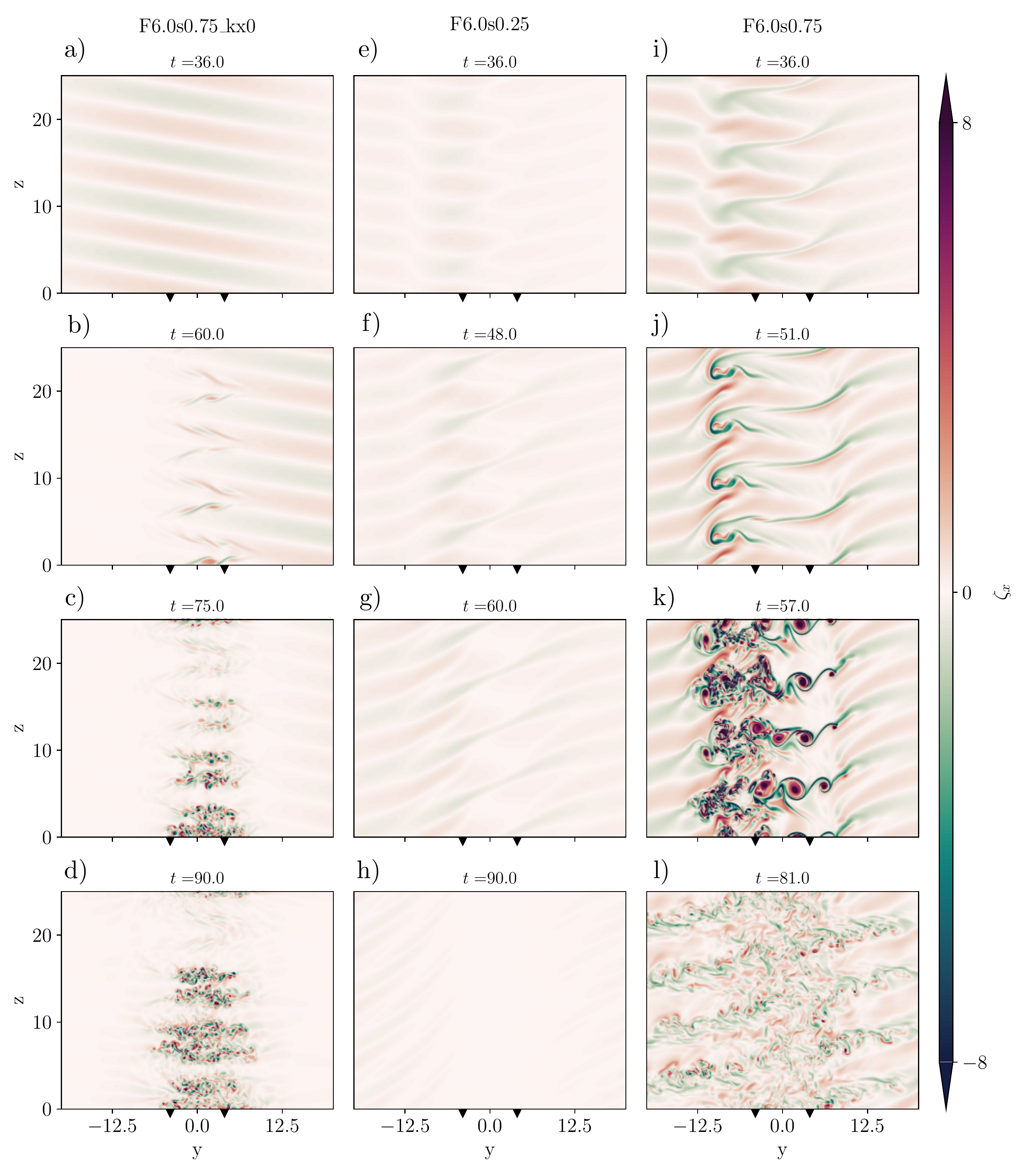}% Here is how to import EPS art
\caption{\label{fig:panels_liftup}  The same as figure 3, for simulations $(a)$-$(d)$ F6.0s0.75\_kx0; $(e)$-$(h)$ F6.0s0.25; $(i)$-$(l)$ F6.0s0.75. Note the difference in colour bar scale.} 
\end{figure}

The time evolution of $\zeta_x$ for three representative simulations in the case $k_x\geq 0$ is shown in figure \ref{fig:panels_liftup} (simulations F3.0s0.25 and F3.0s0.75 are qualitatively similar to F6.0s0.25 and F6.0s0.75). Note that the left and right columns have parameter values corresponding to the linear calculations shown in figure \ref{fig:wkb}$(c)$ and $(d)$. As expected, the dynamics are fundamentally distinct from the wave trapping scenario explored in the previous section. In the special case $k_x=0$ shown in the left-hand column of figure \ref{fig:panels_liftup}, the shear flow does not affect the propagation of the wave so that the slope of phase lines is unchanged in $y$. However, as will shortly be explained, wave advection of mean momentum across the center of the shear layer leads to the local generation of vertical shear. The resulting shear instability creates velocity perturbations which correspond to the small-scale vorticity structures that emerge in the center of the shear layer in panel $(c)$. These local shear instabilities lead to thin layers of turbulence that persist even after the initial wave has passed through the shear layer completely. Curiously, there are noticeable variations in the development of shear instabilities at different vertical heights (most clearly seen around $z\approx 20$). Vorticity perturbations have a sufficiently large vertical extent such that it seems reasonable to suppose that the instabilities developing in one layer influence the dynamics of adjacent layers. We speculate that interactions between layers could quickly magnify initially small differences in the timing of instability onset in neighbouring layers that occur due to random fluctuations in the background noise field.

When $k_x$ is strictly positive, refraction of the incoming wave is observed, though phase lines are now tilted towards the horizontal instead of the vertical direction, as seen in figure \ref{fig:panels_liftup}$(e)$ and $(i)$. Wave reflection caused by the shear leads to the generation of an interference pattern (seen most clearly in panel $e$) between the incident and reflected waves, complicating the subsequent dynamics. Note that the linear theory predicts a turning level at $y=0$ for both cases shown. The flow remains laminar for the smaller amplitude wave scenario; however, for larger initial wave amplitude, instabilities start to develop in the region where incident and reflected waves interact (panel $j$). Panel $(k)$ shows that subsequent instabilities in the form of classical Kelvin-Helmholtz billows develop towards the center of the shear layer, leading to greatly enhanced vorticity and eventually fully developed turbulence.

\begin{figure}
\includegraphics[width=\textwidth]{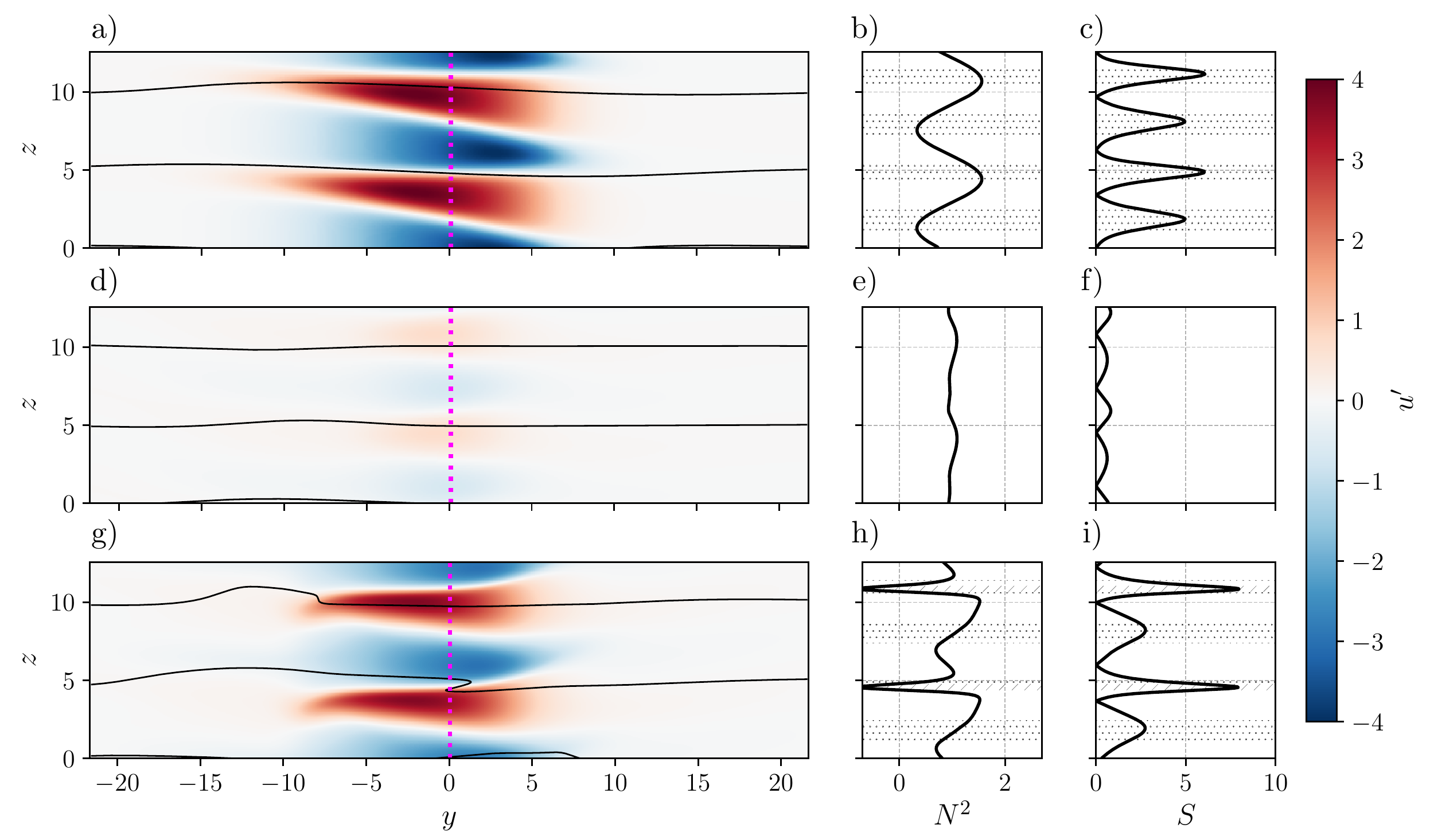}% Here is how to import EPS art
\caption{\label{fig:profiles_liftup}  The same as figure \ref{fig:4}, for simulations $(a)$-$(c)$ F6.0s0.75\_kx0; $(d)$-$(f)$ F6.0s0.25; $(g)$-$(i)$ F6.0s0.75, where snapshots are taken at times corresponding to the top row panels in figure \ref{fig:panels_liftup}. In the right-hand panels, dotted hatching indicates shear-unstable regions where $0\leq Ri_g < 1/4$, and line hatching shows vertical locations where $N^2<0$, as before.}
\end{figure}

The influence of the advection of background shear by the wave is made abundantly clear by plotting the streamwise perturbation velocity $u'$ in figure \ref{fig:profiles_liftup}. Horizontal layers of positive and negative $u'$ are seen in all three cases, with shear magnitudes $O(1)$ and above. In particular, for the large-amplitude initial conditions, $u'$ may be sufficiently large such that regions with $0< Ri_g<0.25$ develop (top and bottom rows), whose locations correspond to the layers of shear instabilities seen developing in figure \ref{fig:panels_liftup}$(c,k)$. This behaviour is in accordance with the ray-tracing predictions of figure \ref{fig:wkb}$(c,d)$. Specifically, the linear theory correctly predicts values of $F$ large enough such that $Ri_g<0.25$ in simulations F6.0s0.75 and F6.0s0.75\_kx0, whereas for simulation F6.0s0.25 (which does not become turbulent), $\min(Ri_g)\approx (1-s_{\mathrm{max}})/F_{\mathrm{max}}^2s_{\mathrm{max}}^2 \approx 0.42 >0.25$. On the other hand, a notable difference from the theory arises from the interference of the incident and reflected waves in simulation F6.0s0.75, leading to narrow regions of local static instability with $N^2<0$ in locations where the instabilities seen in figure \ref{fig:panels_liftup}$(j)$ develop. Clearly in this case, convective instabilities develop on a shorter time scale than shear instabilities in the adjacent regions with $0<Ri_g<0.25$. However, though regions with $N^2<0$ play a dominant role in the triggering of secondary instabilities, it will be shown that the subsequent development of turbulence and mixing remains very much driven by the enhanced vertical shear. 
\begin{figure}
\includegraphics[width=\textwidth]{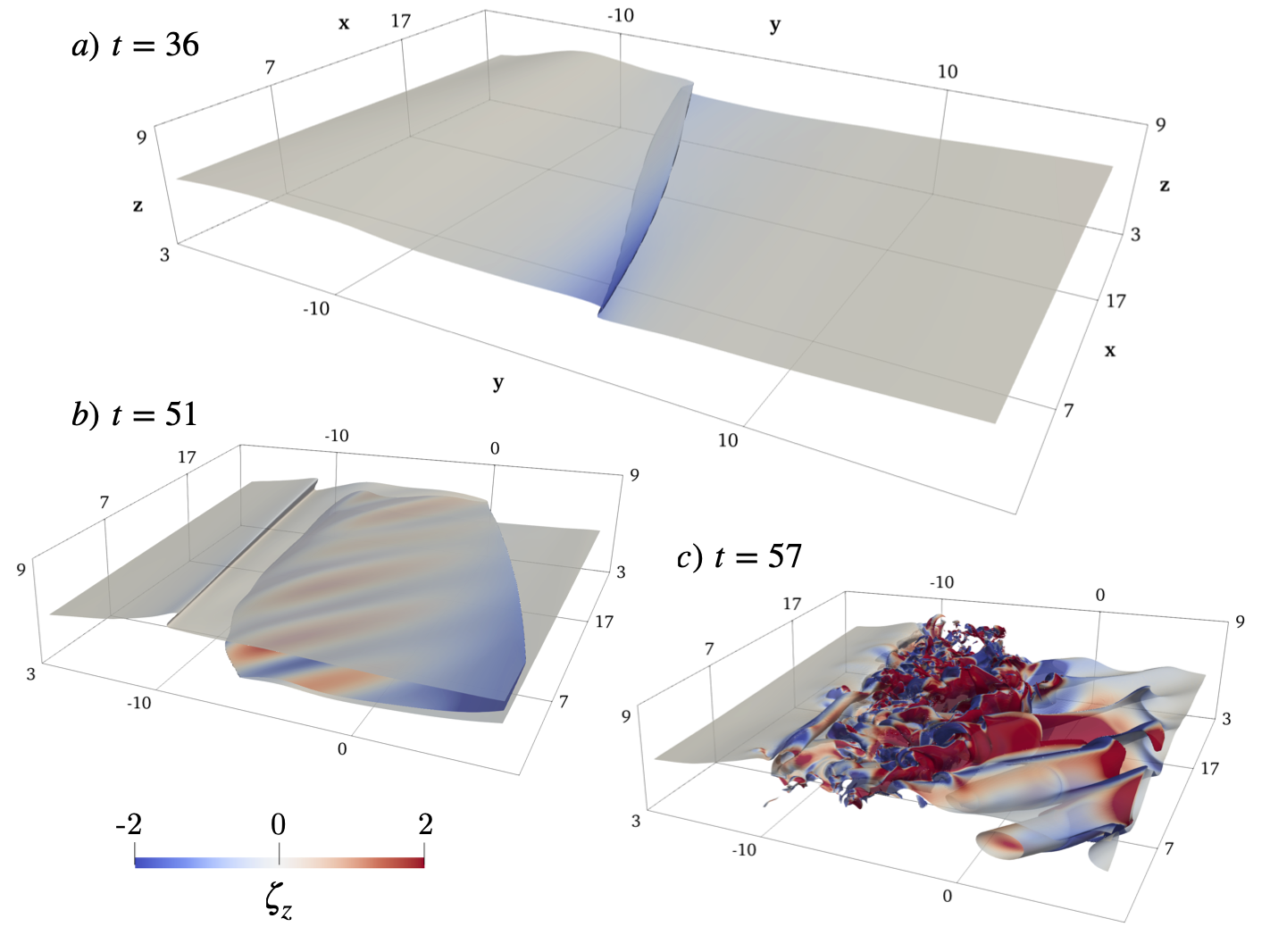}% Here is how to import EPS art
\caption{\label{fig:9} The same as figure 5, for simulation F6.0s0.75.}
\end{figure}
\begin{figure}
\includegraphics[width=\textwidth]{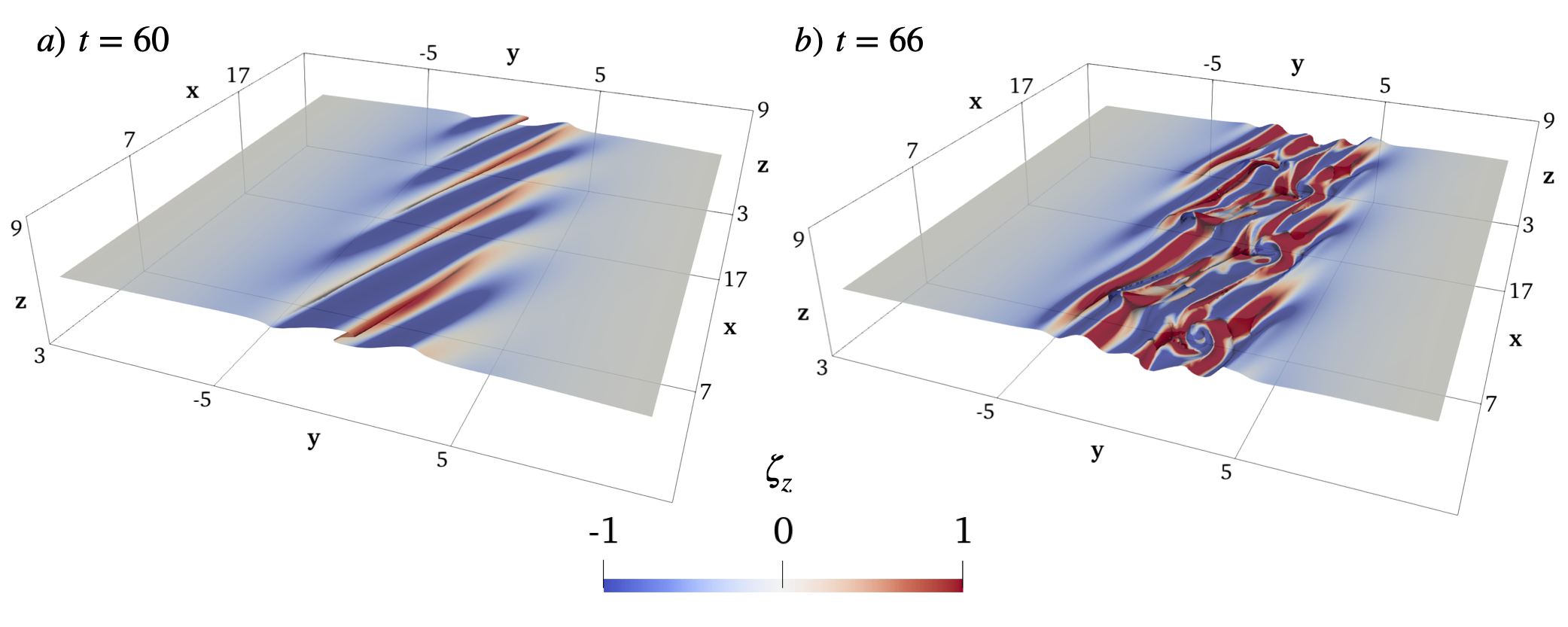}% Here is how to import EPS art
\caption{\label{fig:10} The same as figure 5, for simulation F6.0s0.75\_kx0.}
\end{figure}

A three-dimensional visualisation of simulation F6.0s0.75 is shown in figure \ref{fig:9}. The interaction of the reflected and incident waves at the center of the shear layer leads to the isopycnal overturning seen in panel $(a)$. Panel $(b)$ shows this overturning region spreading out across the shear layer and in particular highlights the wave advection of background flow vertical vorticity that leads to the vertical shear generation. At the same time, streak-like $\zeta_z$ perturbations have emerged in the center of the overturning region. These instabilities evolve to have billow-like structures seen in panel $(c)$ closely resembling those of classical shear instability, where vortices are oriented in the horizontal direction at an oblique angle to the $x$-direction of the background flow. Note that this orientation is not consistent with the prediction $\beta=\pi/2$ from \eqref{eq:localri}, that is, that the most unstable direction is the streamwise direction (and thus the billow axis should be in the $y$-direction). This is likely due to the continuous shearing of the growing perturbations by the background flow. In fact, a similar phenomenon was also reported by \citet{lelong1998inertiaa} for shear instabilities in convectively stable inertia-gravity waves, where the discrepancy between theory and simulations was attributed to the lack of scale separation between the timescales of wave propagation and the growth rate of instability assumed when appealing to a local gradient Richardson number criterion for instability. The development of small-scale turbulent structures seen in panel $(c)$ is due to a combination of convective instability of the overturning region seen in $(b)$ and secondary instabilities of the developing Kelvin-Helmholtz billows. As such, the precise realisation of the turbulent transition is likely to be sensitive to $Re_w$ (and indeed $Pr$). 

Further evidence for behaviour dominated by vertical shear instability is provided by simulation F6.0s0.75\_kx0, which does not exhibit the wave reflection and subsequent isopycnal overturns seen in simulation F6.0s0.75. A three-dimensional picture of developing secondary instabilities for F6.0s0.75\_kx0 is shown in figure \ref{fig:10}. The dynamics are similar to those in figure \ref{fig:9}, with perturbations oriented at an oblique angle to the $x$-direction. In this case, the angle $\beta$ is even further from the predicted value of $\pi/2$, possibly due to the slower growth rate of instability. As in figure \ref{fig:9}, turbulence in figure \ref{fig:10} follows from a subsequent instability of these perturbations. In this case the secondary instabilities seen in panel $(b)$ resemble the instability due to misaligned vortex cores studied by \citet{fritts2022multi}.

\section{Momentum transport}\label{sec:momentum}
\begin{figure}
\includegraphics[width=\textwidth]{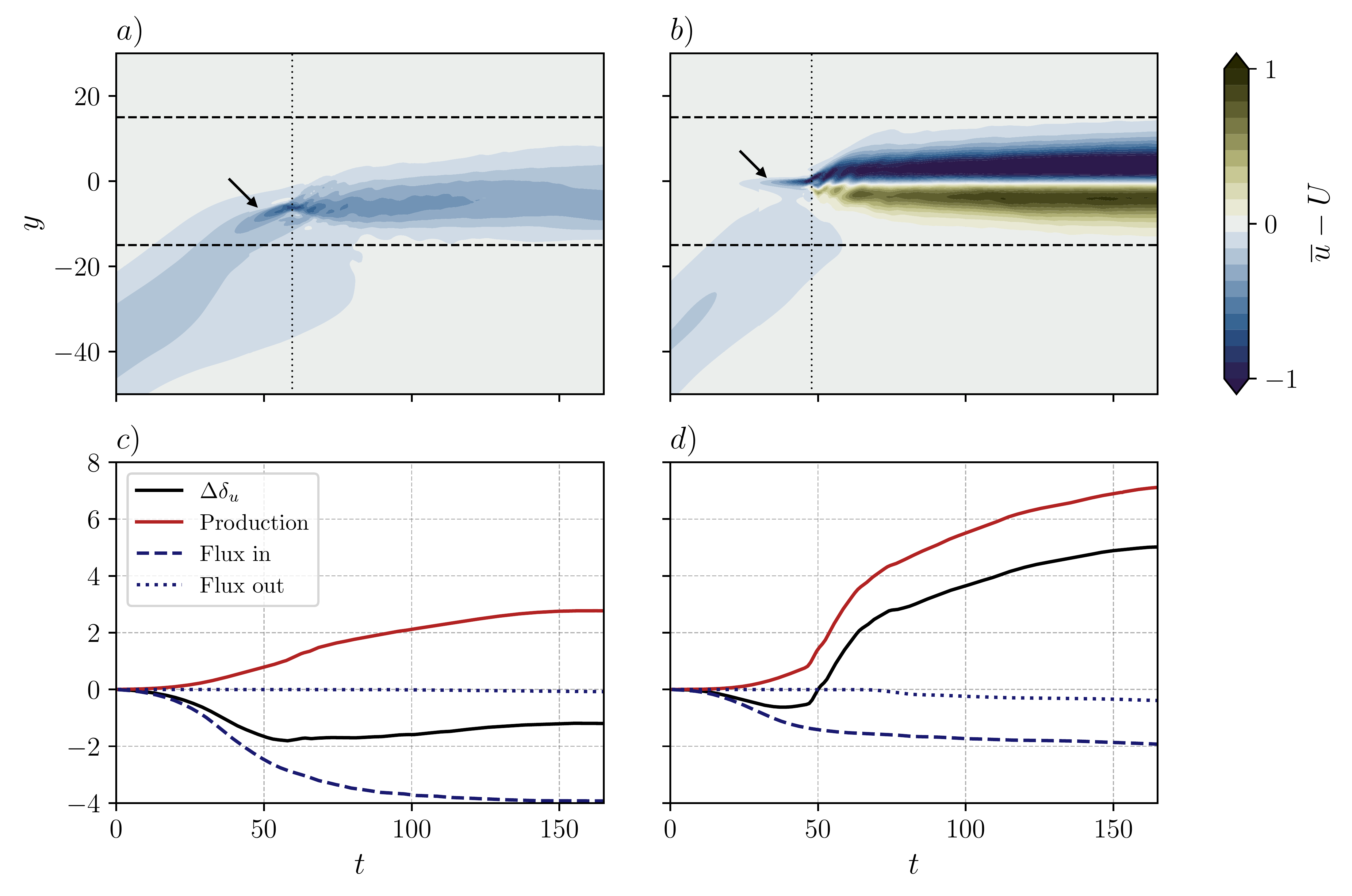}% change x label to t
\caption{\label{fig:momentum_trapped}Top panels show time-evolving, $x,z$-averaged plots of the change in the mean streamwise velocity field from the initial profile $U(y)$, for simulations $(a)$ F0.1s2.28(T), $(b)$ F1.0s5.0(T). Dotted vertical lines indicate the onset of turbulence. Arrows point to the development of a mean flow jet prior to breaking. Bottom panels show corresponding cumulative contributions to the change in the shear layer thickness $\Delta \delta_u$ given by the (time-integrated) terms in equation \eqref{eq:momentum}, averaged over the region indicated by the dashed horizontal lines in the panels above.}
\end{figure}
The evolution of the horizontally averaged background flow $\overline{u}$ is given by
\begin{eqnarray}
    \frac{\partial \overline{u}}{\partial t} = -\frac{\partial }{\partial y}(\overline {u'v' }) + \frac{1}{Re_w}\frac{\partial ^2 \overline{u}}{\partial y^2}.
\end{eqnarray}
%u'v' = kxky s^2/kh^2k^2, cgy = kykz^2/khk^3 -> ubar = kxs^2/(kz^2)/(kh/k) which is exactly the bretherton flow
Note that, neglecting the viscous term and using the wave polarisation relations \eqref{eq:polarisation}, $ \overline{u'v'}$ is initially negative (positive) when $k_x<0$ ($k_x>0$) and the right hand side is thus negative (positive) at the leading edge of the wave packet. It follows that the wave packet carries with it a negative mean flow anomaly when $k_x<0$ and a positive mean flow anomaly when $k_x>0$. This corresponds to a unidirectional Bretherton flow derived in appendix \ref{sec:brethertonflow}. It is also convenient to quantify bulk changes in the momentum of the background flow in terms of the shear layer thickness $\delta_u$ (cf. \citealp*{pham2009dynamics,staquet2002transport}), defined here as
\begin{eqnarray}
    \delta_u = \frac{1}{\Delta u^2}\int_{-15}^{15}\Delta u^2 - \overline{u}^2(y)\, dy. 
\end{eqnarray}
Henceforth, for convenience we denote an integral over $-15<y<15$ with angle brackets $\langle \cdot \rangle$. The evolution of $\delta_u$ is given by 
\begin{eqnarray}
    \frac{1}{2}\frac{d \delta _u}{dt} = -\frac12 \frac{1}{\Delta u^2}\frac{d\langle \overline{u}^2\rangle }{dt} = \left\langle 
    \frac{\partial \overline{u}}{\partial y} \overline{u'v'}\right\rangle -\left[ \overline{u}(\overline{u'v'}) - \frac{1}{Re_w}\overline{u}\frac{\partial \overline{u}}{\partial y}\right]_{-15}^{15} -\frac{1}{Re_w}\left\langle\left(\frac{\partial\overline{u}}{\partial y}\right)^2\right\rangle,\label{eq:momentum}
\end{eqnarray}
where the first and second terms on the right hand side represent the exchange of energy between the mean flow and perturbations (shear production) and the mean momentum flux into and out of the region. The final two terms are due to diffusion of the mean flow and are negligible at the values of $Re_w$ considered here.
% Notice that the component of $u'$ due to wave of advection of momentum is exactly out of phase with $v'$ and thus carries no additional momentum flux divergence. 
\subsection{Cases with $k_x<0$ (wave trapping)}\label{sec:momentumkx>0}
The evolution of $\overline{u}(y)$ relative to the initial shear profile $U(y)$ is shown in figure \ref{fig:momentum_trapped}$(a)$ and $(b)$ for simulations F0.1s2.28(T) and F1.0s5.0(T). The negative anomaly propagating with the wave packet is clearly seen. As the wave packet is refracted and becomes locally steep in panel $(a)$, the anomaly is focussed into a narrow jet and increases in amplitude in a manner closely resembling the self-acceleration mechanism described by \citet{sutherland2006weakly} for internal waves propagating in a vertical plane. For the narrower shear layer in panel $(b)$, a thin jet also develops, though in this case the acceleration does not appear to coincide with a focussing of the wave-induced mean flow but rather emerges within the critical layer located at $y=0$ prior to the arrival of the wave packet center. These dynamics are more similar to the situation described by \citet{wang2020nonlineari} with a steady small-amplitude wave forcing.  The cumulative contribution of the wave-induced mean flow to the shear layer thickness is clear from panels $(c)$ and $(d)$, which reveal the competition between  momentum flux from the wave that acts to decrease the shear layer width and the extraction of momentum by shear production that has the the opposite effect. In both cases, the development of turbulence after the onset of wave breaking results in a drag on the mean flow leading to a reduction in speed. This drag may be significant and in fact overcome the prior acceleration, resulting in the net increase in the shear layer thickness seen in panel $(d)$.
\begin{figure}
\includegraphics[width=\textwidth]{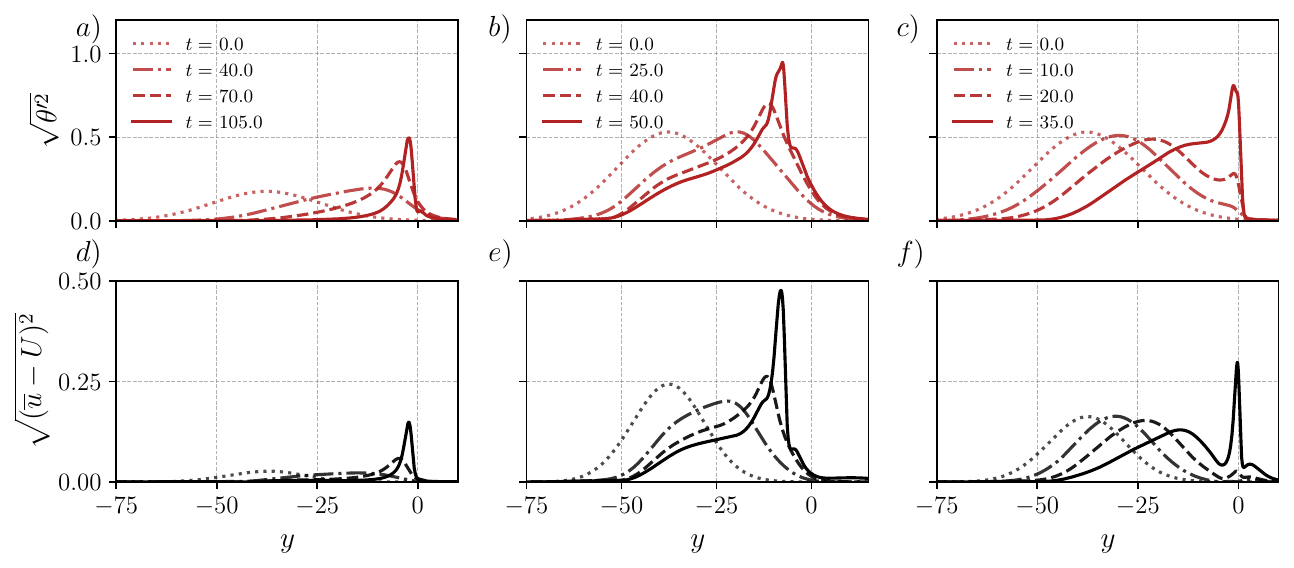}% 
\caption{\label{fig:critical_layer} Horizontally averaged profiles of the root mean square buoyancy perturbation field (top row) and root mean square wave-induced mean streamwise flow (bottom row) at successive time points indicated by the dotted to solid lines. Panels $(a)$ and $(d)$ correspond to simulation F0.1s0.76(T); $(b)$ and $(e)$ to F0.1s2.28(T); $(c)$ and $(f)$ to F1.0s5.0(T). }
\end{figure}
% A notable feature of the linear theory for wave trapping in a horizontal shear flow is the transfer of momentum from the mean flow to wave perturbations. In contrast, it is well known that precisely the opposite occurs when waves are trapped at critical levels in a vertical shear flow. Wave breaking in a horizontal shear layer may thus occur when the steepness $s$ becomes sufficiently large due to the extraction of energy from the background flow, in particular being associated with convective instability when $s>1$. However, the wave-breaking behaviour for simulation F0.1s0.76(T), which does not develop buoyancy overturns, and the unexpected net acceleration of the mean flow observed in simulation F0.1s2.28(T) suggest the picture is more complex than this description suggests. In all cases with $k_x<0$, wave instability occurs in a relatively narrow region and is preceded by a strong acceleration of the mean flow within this region. 
Despite the differences in the manner of formation of the jet seen above, the similarities in its structure and the subsequent wave-breaking suggests underlying similarities in the dynamics driving instability. A key dynamical feature of the nonlinear critical layer studied by \citet{wang2020nonlineari} is the local acceleration of the mean flow by the accumulation of wave pseudo-momentum (see \citealp{buhler2014waves}) within the critical level. Mean profiles of the root-mean-square buoyancy perturbation are shown in figure \ref{fig:critical_layer} alongside the adjustment to the mean flow, which show that formation of the mean flow jet defect is associated with a similarly sharp increase in the local buoyancy perturbation amplitude in all cases. For case F0.1s0.76(T) with a small amplitude, local perturbation and mean flow growth are associated with a focussing of the initial wave energy seen in panel $(a)$. On the other hand, in case F1.0s5.0(T) it is clear from panel $(c)$ that amplification occurs rapidly within the predicted critical layer location significantly prior to the arrival of the center of the wave packet. Despite these differences, the increase in both local mean flow and perturbation energy growth described by \citet{wang2020nonlineari} for a steady wave forcing in unbounded stratified horizontal shear flow appears to be quite generic. %As explored by \citet{wang2020nonlinearii}, an important consideration at the moderate values of $Re_w$ considered here is the role played by viscosity in the critical layer.

\subsection{Cases with $k_x\geq 0$}
\begin{figure}
    \includegraphics[width=\textwidth]{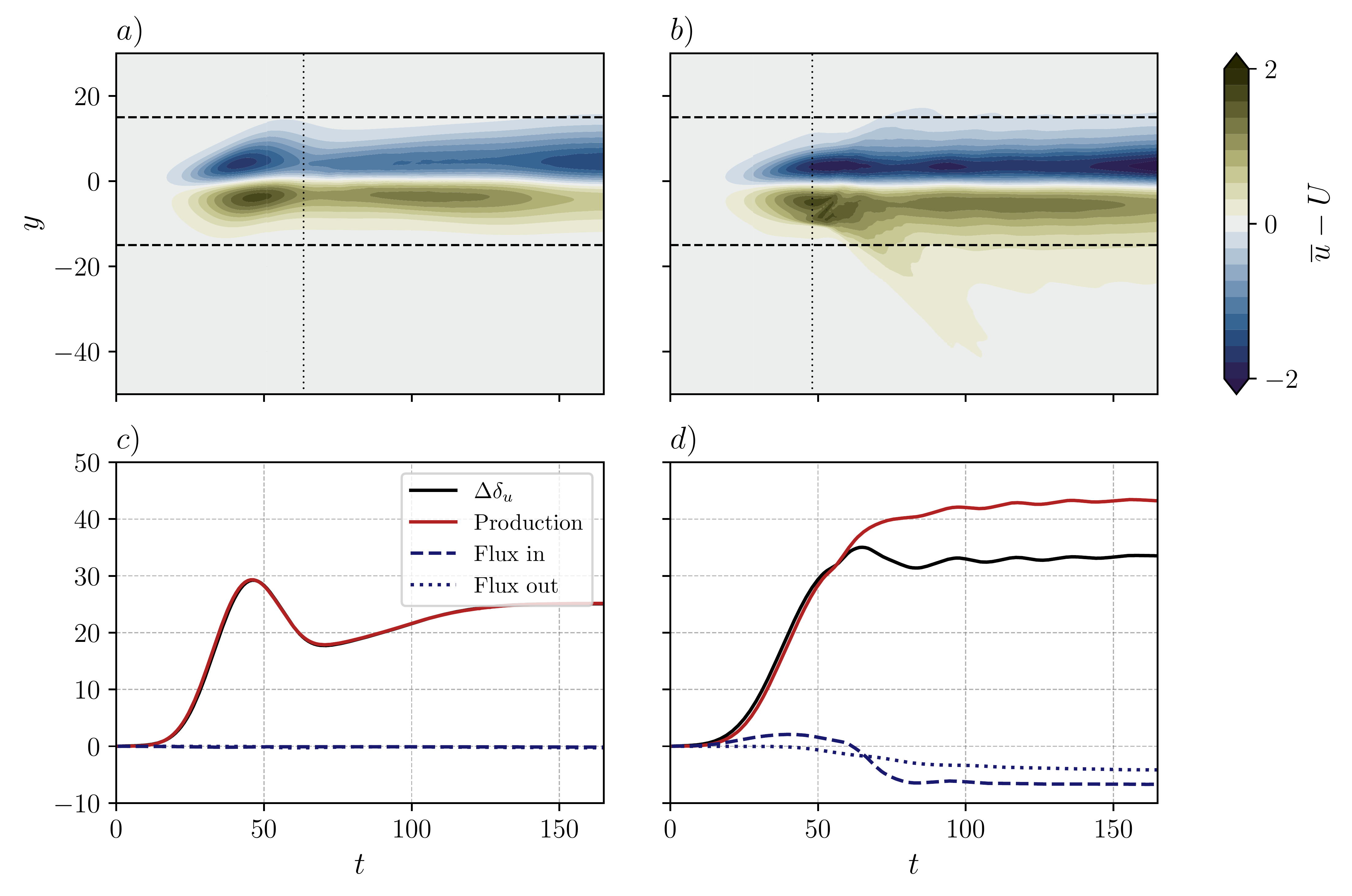}% Here is how to import EPS art
    \caption{\label{fig:momentum_liftup} The same as figure \ref{fig:momentum_trapped}, for simulations F6.0s0.75\_kx0 (left panels) and F6.0s0.75 (right panels).}
    \end{figure}
Figure \ref{fig:momentum_liftup}$(a)$ and $(b)$ show the evolution of $\overline{u}-U(y)$ for cases with $k_x=0$ and $k_x=0.05>0$. The signature of the Bretherton flow in the latter case is not visible due to the comparatively small magnitude of $k_x$. In both cases, significant mean flow deceleration occurs prior to the development of turbulence as momentum is transferred into the streamwise perturbation component, as the local ray-tracing theory suggests. When $k_x=0$ (panel~$a$) and the wave passes through the shear layer, some of this energy is transferred back to the mean flow as the wave starts to exit the shear layer at around $t=50$, before the development of turbulence at around $t=60$. When $k_x=0.05$ and turbulence is energetic, significant wave perturbations radiate outward from the shear layer and decelerate the flow at considerable distances beyond the central region $y=\pm 15$, as seen in panel $(b)$. The asymmetry between the regions to the left and right of the shear layer is due to the reflection of the incident wave. Panel $(c)$ shows that when $k_x=0$, changes to the shear layer thickness are driven entirely by shear production. When $k_x>0$ (panel $d$), the wave-induced mean flow, being proportional to $k_x$, makes a small but positive contribution to the shear layer thickness. The increase in $\delta_u$ is eventually attenuated by mean momentum fluxes out of the domain at both $y=-15$ and $y=15$ due to outward propagating waves, comprised of the reflected incident wave and waves generated by the interaction with the shear.

\section{Energy transport}\label{sec:energy}
% Energy transfers between the incident wave and background flow are diagnosed using a spatial Reynolds decomposition, where a time evolving mean flow $\overline{f}(y,t)$ is defined by letting $\overline{f}$ denote a plane average of the scalar field $f(x,y,z,t)$ over the periodic $x$ and $z$ directions. Perturbations $f'$ are defined according to $f'(x,y,z,t) = f-\overline{f}$. In this way, the mean flow $\overline{\mathbf{u}}(y,t)$ includes the background shear in addition to any adjustments  due to nonlinear wave-mean flow interactions. The perturbation velocity field includes both waves and turbulence. 

We define the total energy density in the region $-15<y<15$ as $E_{tot}= \overline{E}+E'$, where
\begin{eqnarray}
\overline{E} &=& \overline{E}_K+\overline{E}_P=\frac12 \left\langle \overline{u}_i\overline{u}_i\right\rangle  +\frac12  \left\langle \overline{\theta}^2\right\rangle,\\
    E' &=& E'_K +E'_P = \frac12 \left\langle \overline{u_i'u_i'}\right\rangle  +\frac12  \left\langle \overline{\theta'^2}\right\rangle.\label{eq:EKpartition}
\end{eqnarray}
% It is clear from figures \ref{fig:3} and \ref{fig:6} above that the interaction between the incident wave and shear flow involves fluxes of energy due to internal waves both in and out of the shear layer. To quantify these fluxes and separate the dynamics from the wave-shear interaction itself, we integrate energy budget terms over the region $-15<y<15$, denoting this integral with angle brackets $\langle \cdot \rangle$.
The time evolution of the mean and perturbation components $\overline{E}$ and $E'$ is given by
\begin{eqnarray}
    \frac{\partial \overline{E}}{\partial t} &=& -\langle \mathcal{P}\rangle  - \langle \mathcal{D}_{mean}\rangle - [T_{mean}]^{15}_{-15}- [ T_{D,mean}]^{15}_{-15},\\
    \frac{\partial E'}{\partial t} &=& \langle \mathcal{P}\rangle - \langle \mathcal{D}\rangle  - [T]^{15}_{-15} - [T_D]^{15}_{-15}.
\end{eqnarray}
Here, the turbulent production $\mathcal{P}$ represents energy exchanges between the mean flow and perturbations, with dissipation rates for each given by $\mathcal{D}_{mean} $ and $\mathcal{D}$, respectively. Energy fluxes are denoted by $T$ and are separated into diffusive (subscript $D$) and non-diffusive contributions. As in equation~(\ref{eq:momentum}), at sufficiently high $Re_w$ diffusive contributions are typically small. The remaining five terms are defined explicitly as 
\begin{eqnarray}
    \mathcal{P} &=& - \overline{u'v'}\frac{\partial \overline{u}}{\partial y} - \overline{w'v'}\frac{\partial \overline{w}}{\partial y} - \overline{\theta'v'}\frac{\partial \overline{\theta}}{\partial y},\\
    \mathcal{D} &=& \varepsilon' + \chi',\\
    T &=& \overline{p'v'} + \frac12 (\overline{u_i'u_i'v'} + \overline{\theta'\theta'v'}), \label{eq:energytransport}\\
    T_{mean} &=& \frac{1}{2}\overline{u}(\overline{u'v'}) +\frac{1}{2}\overline{w}(\overline{w'v'}),\\
    \mathcal{D}_{mean} &=& \frac{1}{Re_w}\left[\left(\frac{\partial \overline{u}}{\partial y}\right)^2 + \left(\frac{\partial \overline{w}}{\partial y}\right)^2 \right] + \frac{1}{Re_wPr}\left(\frac{\partial\overline{\theta}}{\partial y}\right)^2,
\end{eqnarray}
where $\varepsilon'$ and $\chi'$ are the perturbation dissipation rates of kinetic and (available) potential energy 
\begin{eqnarray}
    \varepsilon ' = \frac{1}{Re_w}\left(\frac{\partial u'_i}{\partial x_j}\frac{\partial u'_i}{\partial x_j}\right), \quad \chi' = \frac{1}{Re_w Pr}\left(\frac{\partial\theta' }{\partial x_j}\frac{\partial\theta' }{\partial x_j}\right).
\end{eqnarray}
% Finally, the rates of change of kinetic and potential energies are given by the equations
% \begin{eqnarray}
%     \frac{\partial E'_K}{\partial t} &=& \mathcal{P} + \mathcal{B} -  \frac{\partial T_K}{\partial y} - \varepsilon',\\
%     \frac{\partial E'_P}{\partial t} &=&  -\mathcal{B} - \chi' - \frac{\partial T_P}{\partial y}.
% \end{eqnarray}
% Here the buoyancy flux  \begin{equation} \mathcal{B}=\overline{\theta'w'}\end{equation} represents the rate of conversion of energy from kinetic to potential energy. 
\begin{figure}
\includegraphics[width=\textwidth]{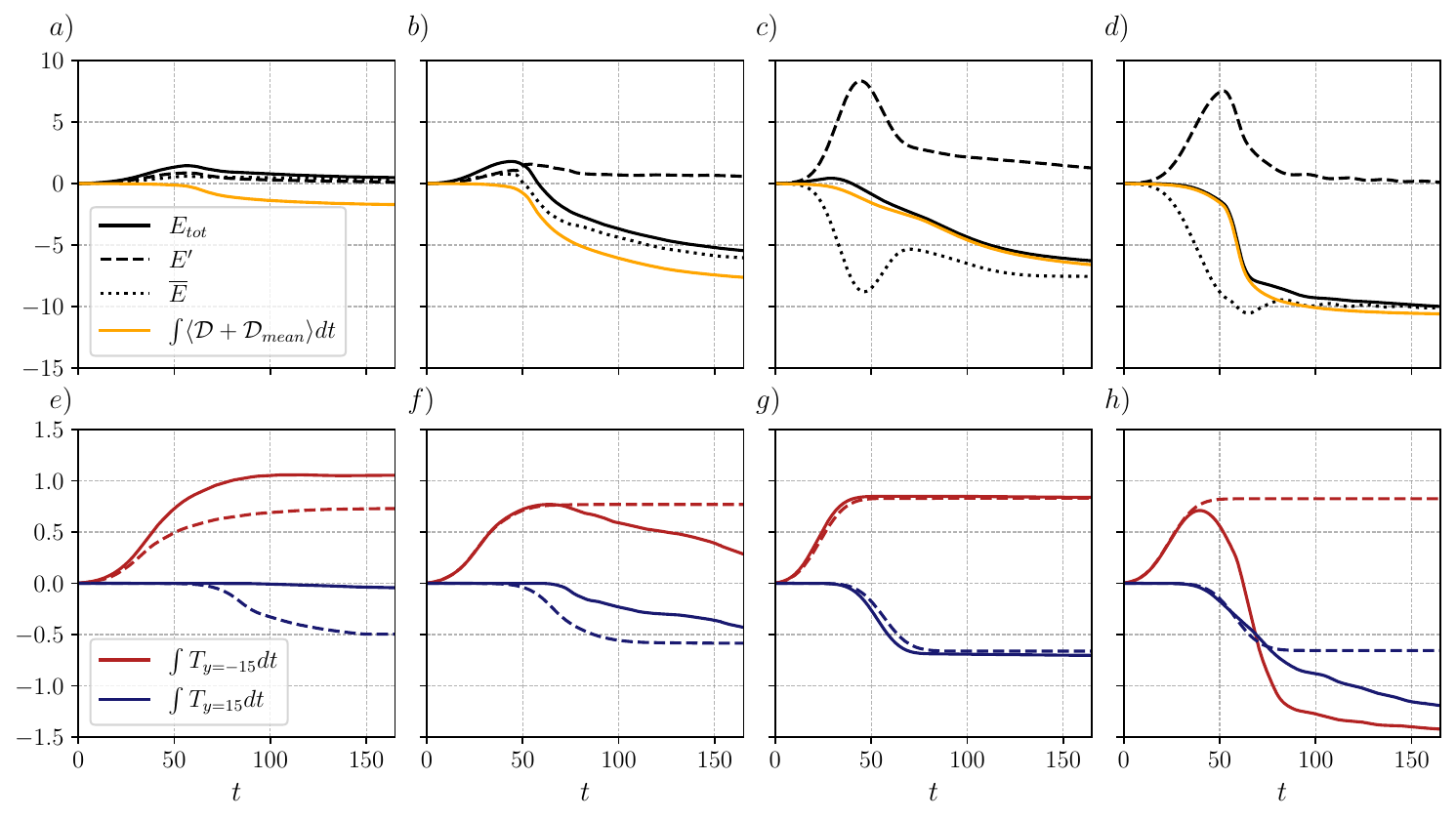}% Here is how to import EPS art
\caption{Top panels show the evolution of the relative total, mean and perturbation energies alongside the cumulative energy dissipation, whilst bottom panels show the cumulative perturbation energy fluxes into and out of the region $-15<y<15$. $(a)$ and $(e)$ are for simulation F0.1s2.28(T), $(b)$, $(f)$ F1.0s5.0(T), $(c)$, $(g)$ F6.0s0.75\_kx0 and $(d)$, $(h)$ F6.0s0.75. Dashed lines in the bottom panels show the cumulative energy fluxes computed from simulations with equivalent initial wave parameters but without the background shear flow. All quantities have been normalised by the initial wave energy. \label{fig:total_energy} }
\end{figure}

The evolution of $ E_{tot} $, $\overline{E} $ and $E'$ within the region $-15<y<15$ is plotted in figure \ref{fig:total_energy} alongside the cumulative energy loss due to viscous dissipation $ \int_0^t \langle \mathcal{D}+ \mathcal{D}_{mean}\rangle dt $. The parameter values for the simulations shown correspond to the cases in figure \ref{fig:wkb}. For simulations F0.1s2.28(T) and F1.0s5.0(T) with $k_x<0$ shown in panels $(a)$ and $(b)$, the total energy increases as the wave propagates into the shear layer, with positive contributions from both mean and perturbation energy. This is consistent with the expected local acceleration due to the wave-induced mean flow described above. The perturbation energy growth reaches a peak at the point where the wave starts to break, which is followed by a rapid loss of mean flow and perturbation energy due to turbulent dissipation. For weaker shear (panel $a$), the cumulative dissipation over the wave-breaking event approaches a value approximately equal to the maximum energy gain due to the wave-shear interaction, which is approximately twice the initial wave energy. However, for stronger shear (panel~$b$), the cumulative dissipation is 7.5~times larger than the initial wave energy. This greatly exceeds the earlier energy growth, suggesting that the turbulent state in this case might more appropriately be thought of as being associated with a subcritical transition of the otherwise linearly stable horizontal shear flow, rather than simply balancing the total increase in energy due to the wave at the point prior to breaking. In particular, the inertial instability suggested by \citet{staquet2002transport} (in which rotation provides a means for additional perturbation energy to be supplied to turbulence) 
%for the same system with added rotation as the means by which additional perturbation energy is supplied to turbulence 
does not appear to be necessary to drive significantly enhanced dissipation. 

The behaviour in simulations F6.0s0.75 and F6.0s0.75\_kx0, shown in figure \ref{fig:total_energy}$(c)$ and $(d)$, is quite different, with wave advection of momentum leading to an immediate and dramatic increase in perturbation energy sourced from the mean flow. Once again, the peak in energy growth corresponds to wave-breaking and the onset of enhanced turbulent dissipation. In simulation F6.0s0.75\_kx0 (panel $c$), the perturbation energy starts to decrease as the wave packet exits the shear layer. While some of this energy is returned to the mean flow, a steady but significant decrease in total energy occurs due to turbulent dissipation associated with vertical shear instability. In contrast, for simulation F6.0s0.75 (panel $d$), there is a much more rapid decrease in the total energy due to enhanced dissipation, indicating a more transient burst of turbulence. The total energy dissipation is significant in both scenarios at between 5 and 10 times the initial wave energy. 

A significant flux of energy out of the region $-15<y<15$ may be expected when the primary wave either passes through or is reflected, or when the wave-shear interaction and/or resulting turbulence excite further waves that radiate outwards. These effects are shown in the bottom panels of figure \ref{fig:total_energy}. To provide reference values for wave transmission, four additional simulations were performed with the same wave parameters but no background shear flow (cf. \citealp{winters1994three}), indicated by the dashed lines in panels $e)$-$h)$. Cumulative energy fluxes through the $y=\pm 15$ boundaries from each simulation where the background shear is present are compared to those from the reference no-shear cases. Note that in all cases the total flux of energy is smaller than the initial wave energy due to losses associated with (non-turbulent) viscous dissipation and dispersive effects, the former of which dominates. In simulations F0.1s2.28(T) and F1.0s5.0(T), wave trapping initially prevents the expected transmission of energy, evident from the solid blue lines in panels $(e)$ and $(f)$. It is notable however that significant generation of internal waves in F1.0s5.0(T) results in outgoing wave radiation, thereby enabling energy transmission. The decrease in the cumulative energy flux is approximately the same through the left and right boundaries suggesting wave generation is roughly symmetric. The cumulative outgoing energy flux (approximately twice the net  flux through $y=15$) is almost equal to the energy of the initial wave packet and, comparing to panel $(b)$, approximately 15\% of the total energy dissipated at $t=165$ (see also figure \ref{fig:energypartition} and the accompanying discussion below). In simulation F6.0s0.75\_kx0 (panel~$g$), the wave energy transmission through the shear layer very closely matches the case with no shear, indicating that despite the significant total dissipation, there is very little energy loss from the incident wave and little to no generation of internal waves by turbulence. In simulation F6.0s0.75 (panel~$h$), the reflection of the initial wave is clear, with a negative cumulative energy flux through $y=-15$. Substantial generation of waves also occurs in this case, adding to the reflected outgoing wave energy and providing significant energy flux through $y=15$. Excluding the reflected incident wave, the total outgoing energy is approximately 20\% of the total dissipation at $t=165$. 

\subsection{Energy partition}
\begin{figure}
\includegraphics[width=\textwidth]{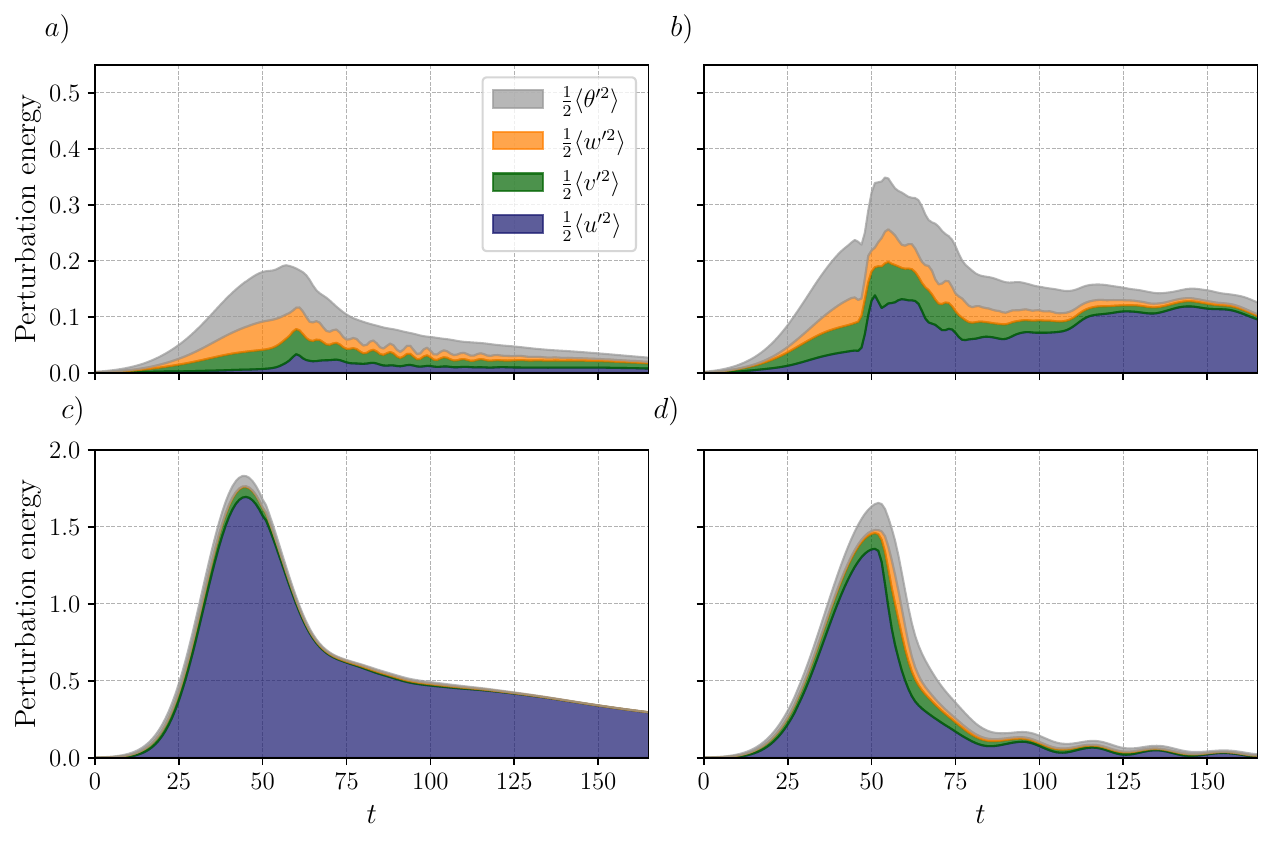}% Here is how to import EPS art
\caption{\label{fig:stacked_energy} Stacked area plots illustrating the split of the total perturbation energy in the region $-15< y<15$ into contributions from individual velocity and buoyancy components for simulations $(a)$ F0.1s2.28(T); $(b)$ F1.0s5.0(T); $(c)$ F6.0s0.75\_kx0 and $(d)$ F6.0s0.75. }
%now need to make this figure correspond to figure 2.
\end{figure}
The ray-tracing predictions in \S\ref{sec:setup} suggest that the interaction between the incident wave and horizontal shear can, at least locally, lead to characteristic differences in the partition between perturbation kinetic and potential energy, and more specifically between the individual components of kinetic energy. Plots of the total perturbation energy $ E' $ and its partition into separate components (equation~\eqref{eq:EKpartition}) are shown in figure \ref{fig:stacked_energy} where, once again, panels correspond to the cases from figure \ref{fig:wkb}. Despite the limitations of the analysis describing local wave energy densities, the bulk energy partition is remarkably consistent with the predictions from the theory. Recall that $F$ represents the excess energy in the streamwise velocity perturbation component $u'$ due to wave advection of momentum, with the maximum instantaneous value $F_{\mathrm{max}}$ from the ray-tracing theory used to characterise simulations in table \ref{tab:simulation-results}. It is immediately clear that perturbation energy growth is dominated by $\frac12 \langle u'^2\rangle$ in panels $(c)$ and $(d)$, for which $F$ is predicted to increase significantly without corresponding growth in the wave steepness (and hence the other components of the total energy). On the other hand, wave steepness growth is seen to dominate proceedings in panels $(a)$ and $(b)$, as is clear from looking at, for example, the larger contribution of the potential energy component $\frac12 \langle \theta'^2\rangle $ to the total perturbation energy. However, as was suggested by the growth of $F$ in figure \ref{fig:wkb}$(b)$ and the snapshots in figure \ref{fig:4}, advection of momentum may still play a significant role in the instability dynamics in such cases. This prediction is further confirmed here by observing that the (maximum) relative magnitude of $\frac12 \langle u'^2\rangle$ is significantly larger in simulation F1.0s5.0(T) with $F_{\mathrm{max}}=1$ than in F0.1s2.28(T) with $F_{\mathrm{max}}=0.1$. It is important to keep in mind that the linear theory only describes the dynamics up to the point of wave breaking (signified by the peaks in perturbation energy in figure \ref{fig:stacked_energy}).

\begin{figure}
\includegraphics[width=\textwidth]{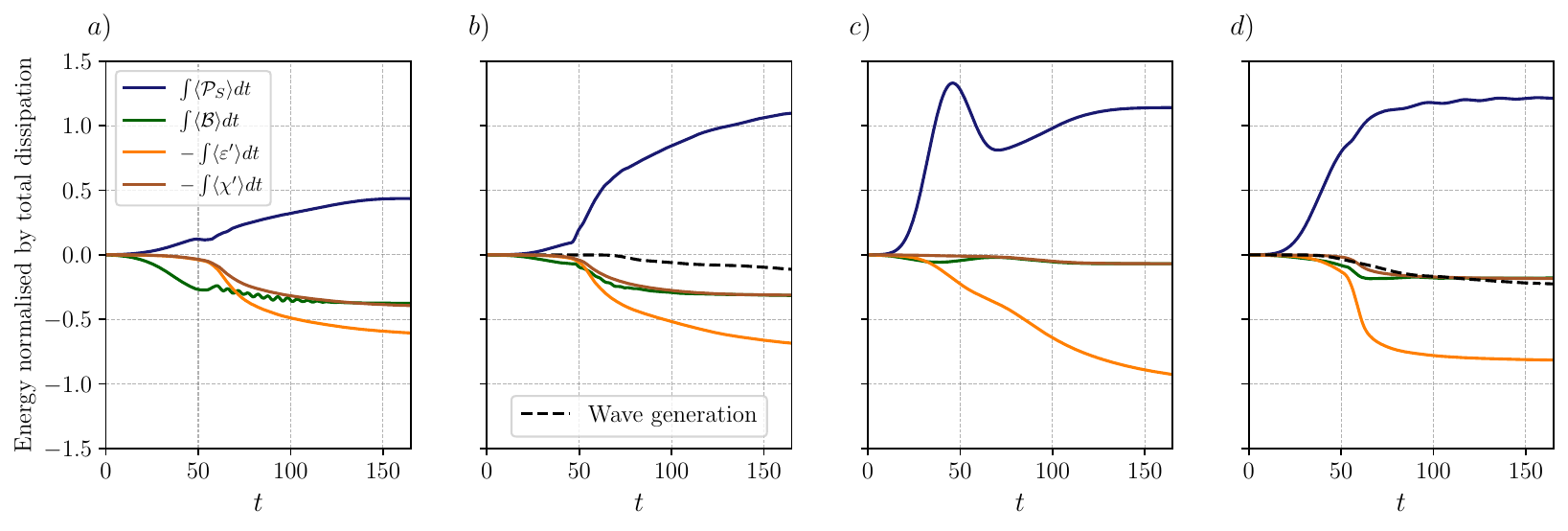}% Here is how to import EPS art
\caption{\label{fig:energypartition}Time-integrated terms from \eqref{eq:EK} and \eqref{eq:EP} for simulations $a)$ F0.1s2.28(T); $b)$ F1.0s5.0(T); $c)$ F6.0s0.75\_kx0 and $d)$ F6.0s0.75. The black dashed lines in panels $b)$ and $d)$, for which there is significant generation of internal waves, show an estimate for the cumulative (total) energy loss due to this emission obtained from doubling the outgoing energy flux through $y=15$ (i.e. assuming symmetric emission). All quantities have been normalised by the cumulative total energy dissipation $\int_0^{165} \langle \mathcal{D} \rangle dt$. }
\end{figure}

A more detailed picture of the exchange of energy between kinetic and potential reservoirs arises from the evolution equations for $E'_K$ and $E'_P$, given by
\begin{eqnarray}
    \frac{\partial E'_K}{\partial t} &=& \langle \mathcal{P_S} \rangle + \langle \mathcal{B}\rangle - \langle \varepsilon'\rangle - \left[\overline{p'v'} + \frac12 (\overline{u_i'u_i'v'})\right]_{-15}^{15} ,\label{eq:EK}\\
    \frac{\partial E'_P}{\partial t} &=&   -\langle \mathcal{B}\rangle -  \langle\chi'\rangle -\left\langle\overline{\theta'v'}\frac{\partial \overline{\theta}}{\partial y}\right\rangle - \left[\frac12(\overline{\theta'\theta'v'})\right]^{15}_{-15},\label{eq:EP}
\end{eqnarray}
where diffusive fluxes have been omitted for brevity. Here the buoyancy flux $ \mathcal{B}\equiv \overline{\theta'w'}$ is the rate of conversion of potential to kinetic energy and the shear production $\mathcal{P}_S\equiv -\overline{u'v'}({\partial \overline{u}}/{\partial y}) - \overline{w'v'}({\partial \overline{w}}/{\partial y})$ is the rate of extraction of kinetic energy from the background flow. The cumulative evolution of these terms is shown in figure \ref{fig:energypartition} alongside the cumulative kinetic and available potential energy dissipation, where terms are normalised by the total cumulative turbulent dissipation late in the simulation at $t=165$. In the wave-trapping cases shown in panels $(a)$ and $(b)$, a significant fraction of the initial energy perturbation growth through the production term is transferred to potential energy via the buoyancy flux. This fraction is slightly reduced in panel $(b)$ when $F$ is larger. On the other hand, panels $(c)$ and $(d)$ confirm the significantly smaller fraction of kinetic to potential energy transfer obvious from figure \ref{fig:stacked_energy}. 

% new paragraph because this is its own thought really
A significant result from figure \ref{fig:energypartition} is that the cumulative dissipation rates of turbulent kinetic and potential energy depend sensitively on the energetics of the wave prior to breaking. Put simply, small-scale turbulent statistics apparently inherit properties of the larger-scale breaking wave. For example, in panels $(c)$ and $(d)$ where wave advection of momentum and hence kinetic energy dominates at the time of breaking, the cumulative dissipation at the end of the mixing event is dominated by $\varepsilon$. On the other hand, when kinetic and potential energy is more equally partitioned at the time of breaking, as in panels $(a)$ and $(b)$, $\chi$ and $\varepsilon$ contribute approximately equally to the total dissipation. Note that $\chi$ is often interpreted as a diapycnal mixing rate (see, e.g. \citealp{howland2021shear}), with a corresponding (cumulative) mixing efficiency given by $\eta(t) \equiv \int_0^t \langle \chi'\rangle dt /\int_0^t \langle \mathcal{D} \rangle dt $. An important observed feature of the horizontally sheared system is that $\eta$ varies considerably between wave breaking scenarios, asymptoting to values from $0.08$ in panel $c)$ to $0.42$ in panel $a)$.
\section{Discussion and conclusions}\label{sec:conclusions}

In this study, we have investigated the problem of an internal gravity wave packet propagating into a horizontal shear layer. When the wave vector forms an oblique angle with the direction of the background flow ($k_x\neq 0$), Doppler shifting leads to local changes in the intrinsic frequency measured by a stationary observer and gradients in the flow velocity refract the wave, leading to changes in the local steepness. When $k_x=0$, wave propagation is unaffected by the shear layer, however significant local perturbation energy changes are possible due to advection of mean flow momentum (equivalent to the lift-up mechanism), as pointed out by \citet{bakas2009gravitya}. 
% The extended WKB theory outlined in this study was shown to account for their combined local effect in a general shear profile $U(y)$. 
In general, the parameter space describing the initial condition is extremely large, thus a reduced order model capturing the essential dynamical variability is desirable. Here, we constructed such a model by studying locally wave-like solutions to the ray-tracing equations under the WKBJ approximation. 
To provide a simple estimate from the initial conditions for the subsequent relative importance of wave advection of momentum, we derived the perturbation energy ratio $F$, where large $F$ corresponds to an excess of kinetic over potential energy due to wave advection of mean flow momentum.
%Where local perturbation growth occurs, a simple estimate from the initial conditions for the subsequent relative importance of wave advection of momentum was shown to be the parameter $F$ that characterises the ratio of kinetic energy to potential energy in the system, where large $F$ corresponds to energy perturbation growth dominated by wave advection of momentum. 

%These two distinct sources of elevated perturbation energy densities were originally identified and explored in the context of transient growth in an unbounded uniform horizontal shear by \citet{bakas2009gravitya,bakas2009gravityb}.

The value of $F$ at a given time during the wave-shear interaction is inversely proportional to the square root of the minimum gradient Richardson number and thus, at least when the latter is positive, serves as an indicator of the susceptibility of the flow to vertical shear instability. Indeed, fully nonlinear DNS with large $F$ and no increase in wave steepness exhibited wave breaking dynamics driven by the emergence of billow-like vortices closely resembling those associated with classical Kelvin-Helmholtz instability (figures \ref{fig:panels_liftup} and \ref{fig:9}). Notably, this picture is also consistent with simulations of plane periodic inertia-gravity waves that, unlike in the non-rotating case, may satisfy $0<Ri_g<0.25$ locally and exhibit similar breaking dynamics to those observed here \citep{lelong1998inertiaa}. On the other hand, when $F$ remains small but wave refraction leads to a significant increase in the local wave steepness, large overturns in the density field are expected play a dominant role in turbulent transition (convective breaking). This prediction is broadly consistent with the dynamics observed in simulations in which waves were trapped by the shear near critical levels, though questions remain regarding the role of the significant local acceleration of the mean streamwise flow in wave breaking and in particular the development of three-dimensional instabilities. The development of this flow appears to be driven by the same fundamental dynamics governing local energy growth in baroclinic critical layers under a steady wave forcing, as described by \citet{wang2020nonlineari}.
\begin{figure}
\centering
\includegraphics[width=0.6\textwidth]{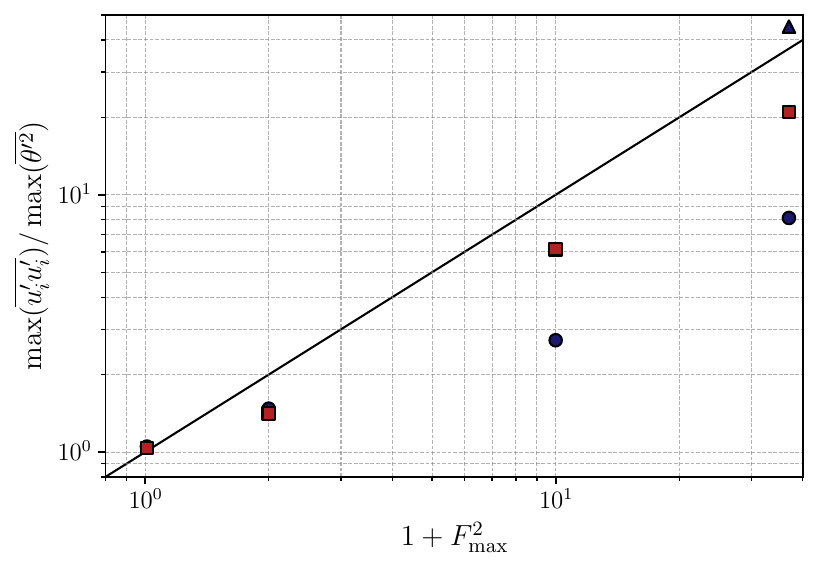}
\caption{\label{fig:16} Predicted ratio of local maximum kinetic to potential energy $1+F_{\mathrm{max}}^2$ from ray-tracing theory versus the value obtained from DNS. Dark blue symbols correspond to simulations with initial wave amplitude $s_0=0.75$ (where a triangle marks the special case $k_x=0$), whilst red symbols have $s_0=0.25$. }
\end{figure}

The effectiveness of the linear theory for describing the local wave state prior to breaking is summarised in figure \ref{fig:16}. In general, there is a strong positive correlation between the DNS data and the predicted local maximum ratio of kinetic to potential energy density. The agreement, though far from exact, is quite satisfactory given the assumptions of linearity and scale separation $k_yh\ll 1$ are generally invalid. At the very least, the use of the theory for describing qualitatively the essential dynamical variability in the wave-mean flow interaction is justified. Unsurprisingly, better agreement is found for simulations with smaller initial wave amplitude. The under-prediction for larger values of $F_{\mathrm{max}}$ when $k_x>0$ in cases where $s_0=0.75$ (compared to smaller values of $s$ and the special case $k_x=0$) is a product of the nonlinear interaction between incident and reflected waves leading to a local increase in the wave amplitude and corresponding overturning. It is worth noting here that the surprising utility of the ray-tracing theory in capturing essential dynamics of wave breaking was also noted for the problem of internal waves in a vertical shear by \citet{howland2021shear}.  

It is important to note that, whilst we have shown that linear ray-tracing theory provides reasonable predictions for the dynamics of instability, energy growth may additionally result from mechanisms not captured under the WKBJ approximation. Firstly, the WKBJ theory is invalid when $k_y h \ll 1$ so that the assumption of scale separation between the waves and the background flow no longer holds. In particular, this occurs near turning levels where $k_y\to 0$. In any case, the theory is invalid even in the linear regime, however, (linear) transient growth may still occur. These dynamics are studied formally in \citet{Arratia2012}, where the formation of layers with strong vertical shear for $k_x\ll 1$ is also observed.  The mechanism for producing these layers, referred to as stratified lift-up, comes from wave advection of mean flow momentum, which also supplies the excess vertical shear predicted in the ray-tracing theory. Secondly, near trapping levels, $k_y\to \infty$, so the scale separation assumption becomes increasingly valid, however, the ray-tracing theory instead becomes invalid due to growing nonlinearities as $s$ increases. Indeed, as discussed in \S\ref{sec:momentumkx>0}, evidence for nonlinear adjustments to the mean flow (through wave self-acceleration or within linear critical levels) has been observed here and warrants attention in future studies.  

In almost all of the DNS, energy perturbation growth eventually led to wave-breaking, turbulent dissipation and mixing. As summarised in figure \ref{fig:total_energy}, the dynamics of turbulent breakdown and the resulting energy exchanges between the mean flow, incident wave and turbulence depend sensitively on the route to wave breaking, and hence on $F$. In both wave trapping and wave reflection scenarios, a large fraction of the total dissipation takes place during a transient burst of turbulence, much like the behaviour observed for wave breaking in a vertical shear flow \citep{winters1994three,howland2021shear}. In notable contrast however, for the horizontal shear flow studied here, larger values of $F$ are associated with a total dissipation that may be several times larger than the energy of the primary wave, thus requiring a significant input of energy to turbulence from the background flow, whose linear stability is enforced here. The corresponding horizontal mixing of momentum results in significant deceleration of the background shear flow, as discussed in \S\ref{sec:momentum}. This may be compared with \citet{winters1994three} for example, who found that the total dissipation is only around 20-25\% of the initial wave energy in a vertical shear with minimum gradient Richardson number equal to $0.5$.

In some cases, the wave-shear interaction led to the generation of internal waves whose total energy was seen to be up to 25\% of the total energy dissipated by turbulence and more than twice that contained in the initial wave packet (figure \ref{fig:energypartition}). We note that even in the absence of turbulent breaking, strong generation of internal waves for $F=O(1)$ was predicted by \citet{bakas2009gravityb}, who found that up to 30\% of the maximum wave energy within the shear layer may be radiated away. On the other hand, localised decaying initially homogeneous and isotropic turbulence can also produce significant internal wave emission in a background stratification. Using a combination of laboratory experiments and direct numerical simulations, \citet{maffioli2014evolution} found the total outgoing wave energy to be 13\% of the cumulative dissipation for a horizontally localised patch of turbulence. To our knowledge, the substantial emission of internal waves by turbulence has not been reported for wave breaking at a vertical critical level (this being most obviously differentiated from the often-observed phenomenon of partial reflection by the symmetry of wave emission between each side the critical level). However, for turbulence produced in an unstable vertical shear layer in the presence of a strongly stratified region below, we can infer from \citet{pham2009dynamics} that internal wave emission constituted up to 22\% of the total energy dissipated. 

An analysis of the turbulent kinetic and potential energy budgets demonstrated that the exchanges between (perturbation) kinetic and potential energy in the lead up to wave breaking had a substantial influence on the subsequent turbulent state, most notably being closely related to the cumulative dissipation rates of kinetic and available potential energy (figure \ref{fig:energypartition}). As a result, waves that broke more convectively, here characterised by approximate equipartition between turbulent kinetic and potential energy at the point of breaking, exhibited corresponding values of the (cumulative) mixing efficiency up to $0.42$, considerably larger than the typical value of $0.17$ assumed in the literature. When instability was driven by vertical shear, mixing efficiencies as low as $0.08$ were observed. In this study, a relatively small number of representative simulations were studied in detail to highlight the influence of the breaking mechanisms on the subsequent turbulent dynamics. The results naturally motivate further simulations spanning a wider range of (at least) $F$, $\Delta u/h$ and $Re_w$ to investigate in more detail the mixing variability and its dependence on turbulent parameters in addition to the mechanisms driven wave breaking. 

Nonetheless, the close relationship between the partition of kinetic and potential energy (which is dominated by larger scale motions) and mixing efficiency is generally consistent with results obtained from strongly stratified forced systems for which turbulence production is maintained against dissipation in a statistically stationary state by continual power input at large scales (see for example \citealp{maffioli2016mixing}). However, the precise relationship to the mixing efficiency is sensitive to other dimensionless turbulent parameters and, importantly, the type of forcing used. Despite broad similarities in the anisotropic structure of turbulence and the shapes of turbulent kinetic and potential energy spectra, considerable differences in mixing properties are observed between forcing concentrated in vertically rotational modes with $k_z=0$, forcing by internal waves and forcing with a mean vertical shear \citep*{lindborg2007stratified, howland2020mixing,portwood2019asymptotic}. The results of this study indicate a wide range of possible local transient wave-breaking dynamics may contribute towards the considerable mixing variability seen in forced systems, which could in theory collectively comprise many such local wave-mean flow interactions depending on the type of forcing. Further investigation is required in order to to link robustly the local properties of wave breaking studied here to their collective influence in complex multiscale systems, where the distinction between the largest scales of turbulence and the smallest scales of internal waves is unclear. 

Throughout this study, we have neglected the influence of planetary rotation that will undoubtedly be important when the time scales of either the shear layer or the internal wave are sufficiently long, that is, when $\Omega\sim O(f)$ or $\partial U/\partial y\sim O(f)$. The latter condition was studied for $N\sim f$ by \citet{staquet2002transport}, who suggested that in the case of wave trapping, inertial instability may act as an additional source of turbulence energy. The former condition may be particularly relevant for scenarios where wave advection of momentum dominates energy perturbation growth, which is favoured for small horizontal wavenumbers $k_y,k_x \ll 1$ and hence small $\Omega/N$. In this case, the dynamics in the presence of rotation may be considerably more complicated, as is suggested by the recent results of \citet*{hilditch2025}. The effect of changing the Reynolds number $Re_w$ will undoubtedly be important during wave breaking and the turbulence that follows, especially at the modest values considered here. In particular, the extents to which the critical layer dynamics (e.g. \citealp{wang2020nonlinearii}), secondary instabilities (e.g. \citealp{Mashayeka13})  and the dissipation and mixing rates shown in figure \ref{fig:energypartition} are affected by changes in $Re_w$ remain open problems. A further parameter that continues to present a considerable computational challenge in direct numerical simulations is the Prandtl number $Pr$ that we have assumed to be unity, but in reality is expected to be $O(10)$ in thermally-stratified ocean environments.

Finally, to what extent the variability in wave breaking dynamics explored here extends to a more general class of background flows--which might include cyclostrophically balanced flows or, for non-zero $f$, rotationally balanced flows (see, e.g. \citealp{whitt2013near,edwards2005focusing})--remains an open question. Scenarios such as these, in addition to the problem studied here, may help to provide a crucial link between wave breaking and the asymptotic regime of anistropic stratified turbulence that has been proposed as a universal description for the dynamics of turbulent patches in the ocean interior \citep{dasaro2022internal,kunze2019unified,riley2008stratified}. As \citet{dasaro2022internal} points out, the initial condition for the classical problem of a (horizontally infinite) wave packet incident on a vertical shear flow lacks potential vorticity, thought to be a key ingredient in the energy cascade of strongly stratified turbulent flows \citep{lindborg2006energy}. The inclusion of a horizontally varying mean flow provides a possible source of potential vorticity at large scales. It would be worthwhile to investigate whether the turbulence produced in the wave breaking scenarios studied here can access this strongly-stratified regime in the case when the mean flow is linearly stable (the unstable case is discussed in detail in \citet{lewin2024evidence}). It is also worth recognising that, in general (and as was the case in this study), horizontally bounded wave packets carry with them their own source of potential vorticity in the form of a Lagrangian vorticity dipole \citep{buhler2005wave, bretherton1969mean}. This may, as suggested by \citet{dasaro2022internal} and by the focussing of the mean flow shown in figure \ref{fig:critical_layer}, play an important role in the resulting energy cascade to small scales during breaking. 

\backsection[Acknowledgements]{This work was initiated at the Stanford Center for Turbulence Research Summer Program 2024. We thank the organisers and participants for their efforts in facilitating an engaging and stimulating research environment. Discussions with Leif Thomas, Jeff Koseff and Bruce Sutherland are gratefully acknowledged.}

\backsection[Funding]{MMPC acknowledges support from the Natural Sciences and Engineering Research Council of Canada (NSERC) through RGPIN-2024-06184. SFL and AKK were supported by the Ho-Shang and Mei-Li Lee Faculty Fellowship in Mechanical Engineering at UC Berkeley. AB acknowledges support from a National Science Foundation Graduate Research Fellowship (Grant Number DGE-1656518). This research used the Savio computational cluster resource provided by the Berkeley Research Computing program at the University of California, Berkeley (supported by the UC Berkeley Chancellor, Vice Chancellor for Research, and Chief Information Officer).} 
% Please provide details of the sources of financial support for all authors, including grant numbers. Where no specific funding has been provided for research, please provide the following statement: "This research received no specific grant from any funding agency, commercial or not-for-profit sectors." }

\backsection[Declaration of interests]{ The authors report no conflict of interest.}

% \backsection[Data availability statement]{The data that support the findings of this study are openly available in [repository name] at http://doi.org/[doi], reference number [reference number]. See JFM's \href{https://www.cambridge.org/core/journals/journal-of-fluid-mechanics/information/journal-policies/research-transparency}{research transparency policy} for more information}

\backsection[Author ORCIDs]{ S. Lewin, https://orcid.org/0000-0002-2602-4751; A. Kaminski, https://orcid.org/0000-0002-4838-2453; A. Balakrishna, https://orcid.org/0009-0006-2696-3130; M. Couchman, https://orcid.org/0000-0002-4667-6829}

% \backsection[Author contributions]{Authors may include details of the contributions made by each author to the manuscript'}

\appendix

\section{Minimum gradient Richardson number}\label{sec:appa}
After some algebra, equation \eqref{eq:localri} for the gradient Richardson number can be written as 
\begin{eqnarray}
    Ri_g =  \frac{R^2}{F^2s^2}\frac{1+s\cos\phi}{\left[\left(R\cos\phi + (k_x/k_y)\sin\phi\right)^2  + \sin^2\phi\right]\sin^2(\beta+\gamma)},
\end{eqnarray}
where $\tan \gamma = \sin\phi/(R\cos\phi + k_x\sin\phi/k_y)$. Thus
\begin{eqnarray}\label{eq:localri_simp}
    \min_{\phi, \beta} (Ri_g) = \frac{R^2}{F^2s^2}\min_{\phi} \left(\frac{1+s\cos\phi}{\left(R\cos\phi +(k_x/k_y)\sin\phi\right)^2+\sin^2\phi}\right).
\end{eqnarray}
In the following we will assume $s<1$ so that $Ri_g>0$. Consider first the case $k_x=0$. For $R\geq 1$, the denominator and numerator of the function in brackets are respectively maximised and minimised at the same value of $\phi=\pi$, making this a global minimiser. For $R<1$, let $c=\cos\phi$, then stationary points of the function in brackets above are given by the roots of a quadratic equation:
\begin{eqnarray}
    c_{min}=\frac{1}{s}\left(-1 \pm \sqrt{1-s^2/(1-R^2)}\right).
\end{eqnarray}
There are two cases to consider. When $1-R^2<s^2$, the discriminant in the expression above is negative and there are no stationary points in the range $-1<c<1$. The global minimum thus occurs at $c=-1 \iff \phi=\pi$. On the other hand, when $1-R^2>s^2$ there is a single global minimum in the range $-1<c<1$ at $c=c_{min}$. Substituting in we obtain:
\begin{equation}
\min(Ri_g)=
\begin{cases}
\dfrac{R^2}{2F^2s^2}
\left(
\sqrt{1+\dfrac{s^2}{R^2-1}}+1
\right),
& R^2\leq 1-s^2,\\[2ex]
\dfrac{1-s}{F^2s^2},
& R^2> 1-s^2.\\
\end{cases}
\end{equation}
Note in particular that $R^2/F^2>1 = $ so that, when $R^2\leq 1-s^2$, $\min(Ri_g)>1/2$ for $s<1$. 
% Since our simplified model of instability requires $s>1$ or $Ri_g<1/4$, cases of interest occur when $(1-s)/F^2s^2<1/2$. 

\begin{figure}
\centering
\includegraphics[width=\textwidth]{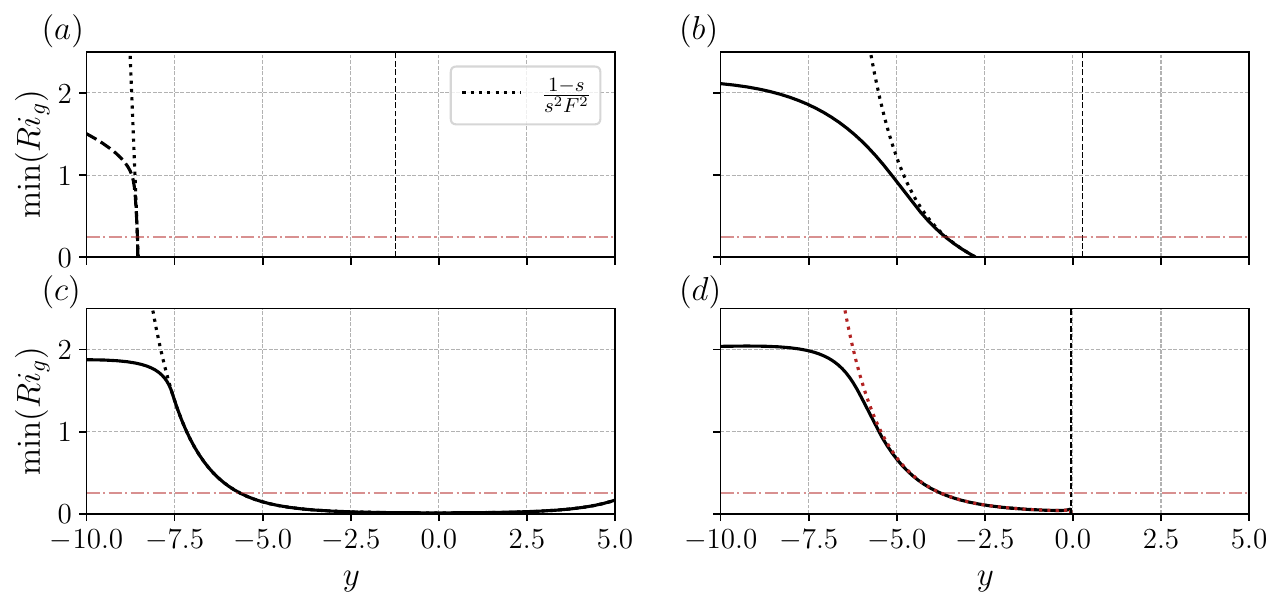}
\caption{\label{fig:rigmin}  Evolution of the minimum gradient Richardson number along rays for the parameter values considered in figure \ref{fig:wkb}, computed by evaluating \eqref{eq:localri_simp} numerically. Dotted curves show the estimate $\min(Ri_g)=(1-s)/F^2s^2$. The heuristic local stability threshold $\min(Ri_g)=1/4$ is also indicated. }
\end{figure}

When $k_x\neq 0$, the denominator of \eqref{eq:localri_simp} is not maximised at $\phi=\pi/2$ for any value of $R$. However, the value at $\phi=\pi/2$ will be similar to the maximum value when $Rk_y/k_x $ is large. Because $Rk_y/k_x = F k_h |\mathbf{k}|/(k_xk_z)>F$, we can argue heuristically that this may be anticipated near locations of wave instability. Where dynamics are dominated by wave steepening near trapping levels, $k_h|\mathbf{k}|/(k_xk_z) \gg 1$. On the other hand, where vertical shear produced by wave advection of momentum dominates, $F$ is necessarily large. Figure \ref{fig:rigmin} demonstrates that, indeed, $(1-s)/F^2s^2$ becomes an increasingly good approximation for $\min (Ri_g)$ as the latter decreases towards zero. Note that, in particular, the point at which $\min(Ri_g)$ becomes smaller than $1/4$ is very well estimated using $(1-s)/F^2s^2$ in all cases.
\section{Inclined coordinates}\label{sec:appb}
\begin{figure}
\centering
\includegraphics[width=0.6\textwidth]{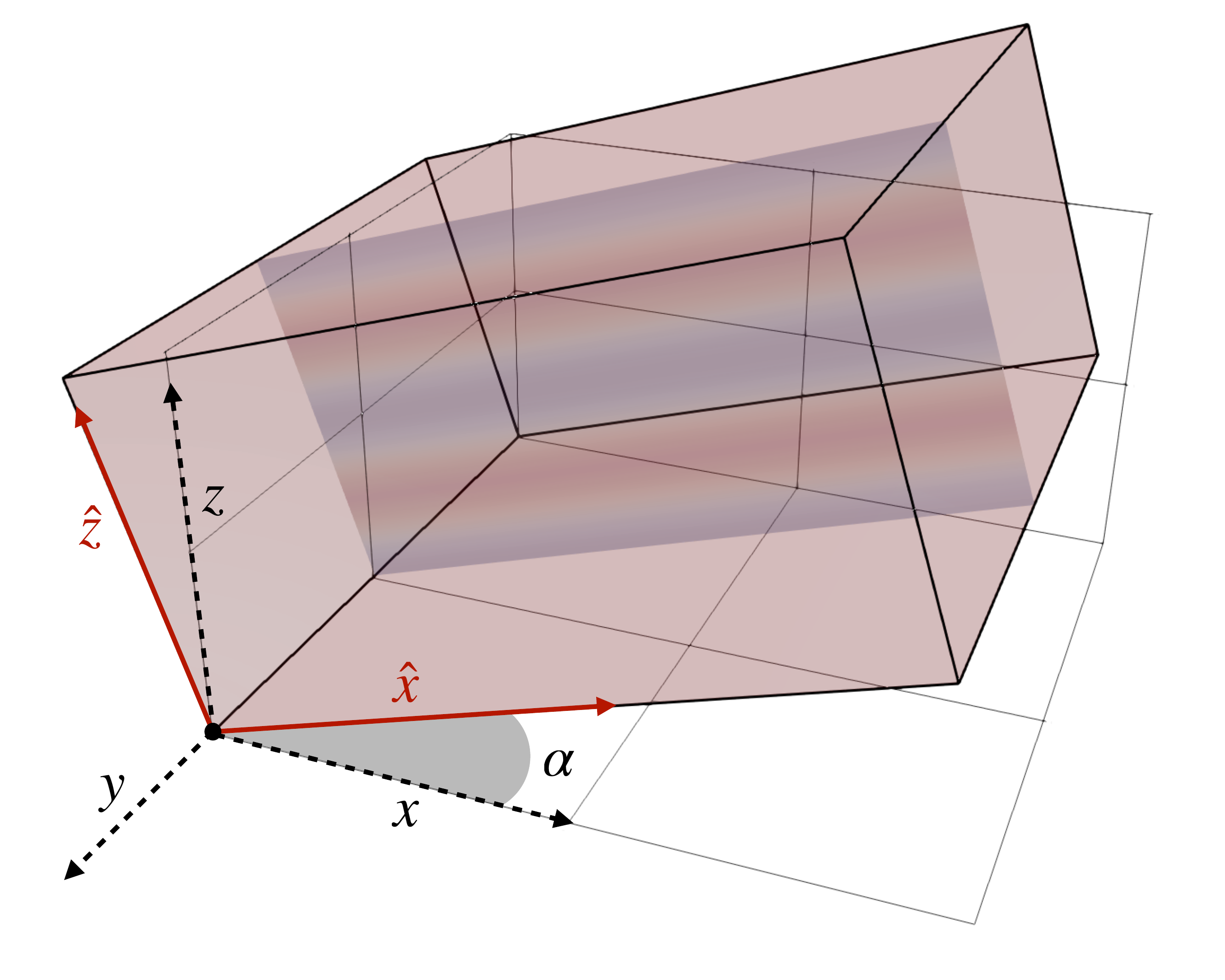}
\caption{\label{fig:rotation} Schematic illustrating the coordinate transformation used in simulations. A translucent slice of the initial vertical velocity field is shown to demonstrate that wave phase lines in the $\hat{x}$$\hat{z}$ plane are parallel to the $\hat{x}$-direction. }
\end{figure}
A schematic of the inclined coordinate system introduced in section \S\ref{sec:incline} is shown in figure \ref{fig:rotation}. Under rotation by an angle $\alpha = -\arctan(k_{x0}/k_{z0})$ in the $x$-$z$ plane, the velocities in the $\tilde{x}$ and $\tilde{z}$ directions are $(\hat{u}, \hat{w}) = (u\cos\alpha + w\sin\alpha, -u\sin\alpha+w\cos\alpha$). The equations of motion are 
\begin{subequations}
\begin{eqnarray}
    &\displaystyle \frac{\mathrm{D}\hat{\mathbf{u}}}{\mathrm{D}t} = \nabla \hat{p} + \hat{\theta}(\hat{\mathbf{e}}_z\cos\alpha + \hat{\mathbf{e}}_x\sin\alpha) + \frac{1}{Re_w} \nabla^2\hat{\mathbf{u}},& \\
    &\displaystyle \nabla \cdot \hat{\mathbf{u}} = 0,& \\
    &\displaystyle \frac{\mathrm{D}\hat{\theta}}{\mathrm{D}t} = -\hat{w}\cos\alpha - \hat{u}\sin\alpha + \frac{1}{Re_wPr}\nabla^2 \hat{\theta},& 
\end{eqnarray}
\end{subequations}
% It is straightforward to check that $k_{x0} x + k_{z0}z = \tilde{k}_{z0}\tilde{z} = k_{z0} \tilde{z}/ \cos\phi$, thus the initial conditions from equation (5) are independent of $\tilde{x}$.
with initial conditions
\begin{subequations}
\begin{eqnarray}
    &\hat{\mathbf{u}}(\hat{\mathbf{x}},t=0) = \Delta u \cos\alpha \tanh(y/h) \hat{\mathbf{e}}_x - \Delta u \sin\alpha \tanh(y/h) \hat{\mathbf{e}}_z +\hat{\mathbf{u}}_\mathrm{wave},  & \\
    &\hat{\theta}(\hat{\mathbf{x}},t=0)=\hat{\theta}_\mathrm{wave},&
\end{eqnarray}
\end{subequations}
where $\hat{\mathbf{u}}_\mathrm{wave}$ and $\hat{\theta}_\mathrm{wave}$ are the plane wave solutions from (5) transformed into the inclined reference frame and expressed in terms of $\hat{\mathbf{x}}$.

\section{Wave-induced mean flow}\label{sec:brethertonflow}
As originally demonstrated by \citet{bretherton1969mean}, the stress exerted by an internal wave packet on the fluid through which it propagates induces a mean flow at second order in the wave amplitude. The nature of this mean flow depends on the geometry of the wave packet, in particular the aspect ratio of its envelope. \citet{van2018wave} provide a general expression for the wave-induced mean flow in a wave packet with arbitrary amplitude envelope $A(x,y,z,t)$, writing the result as the sum of the horizontal Bretherton flow forced by vertical vorticity and a long wave flow forced by horizontal vorticity. Because the wave packets we consider here are unbounded in the vertical, the long wave flow is zero at leading order. Extending the analysis of \citet{van2018wave} to a 3D wave packet, horizontal divergence of the momentum flux drives a mean flow $\mathbf{u}_{DF}$ given by 
\begin{eqnarray}
    \mathbf{u}_{DF} = u_{DF}\left(1,\frac{c_{gy}}{c_{gx}},\frac{c_{gz}}{c_{gx}}\right),
\end{eqnarray}
where $\mathbf{c}_g = (c_{gx}, c_{gy}, c_{gz})$ is the linear group velocity of the waves and 
\begin{eqnarray}
    u_{DF} =\frac{k_x}{\omega} \frac{N^2}{2k_z^2}s^2A^2,
\end{eqnarray}
taking the form of the pseudomomentum per unit mass in the $x$-direction. This so-called divergent-flux induced flow must induce a secondary response flow to ensure incompressibility; the sum of these two flows gives the Bretherton flow $\mathbf{u}_{BF}$. As shown in \citet{van2018wave}, $w_{BF}=0$ and we must have additionally that $(\nabla \times \mathbf{u}_{DF})\cdot \hat{\mathbf{k}} = (\nabla \times \mathbf{u}_{BF})\cdot \hat{\mathbf{k}}$. In combination with the incompressibility condition $\nabla\cdot \mathbf{u}_{BF}=0$, it follows that, in Fourier space with coordinates $(\kappa,\lambda,\mu)$,
\begin{eqnarray}
    \hat{\mathbf{u}}_{BF}= \hat{u}_{DF}\begin{bmatrix}
  [\lambda^2 - (k_y/k_x)\mu \kappa]/\kappa_H^2\\
  [(k_y/k_x)\kappa^2 - \kappa\lambda]/\kappa_H^2 \\
  0
\end{bmatrix},\label{eq:ubf_fourier}
\end{eqnarray}
where a hat denotes the 3D Fourier transform and $\kappa_H^2 = \kappa^2+\lambda^2$. Note that, with $A \sim \exp(-y^2/a^2)$, we have 
$\hat{u}_{DF} = k_xN^2/(2\omega k_z^2)\mathcal{A}(\lambda)\delta(\kappa)\delta(\mu)$, where $\mathcal{A}(\lambda) = \frac{1}{2\pi}\int_{-\infty}^{\infty} A^2(y)\exp(-i\lambda y)dy$ is the Fourier transform of $A^2(y)$ in the y-direction. Substituting this into \eqref{eq:ubf_fourier}, it is straightforward to inverse Fourier transform and find that 
\begin{eqnarray}
    \mathbf{u}_{BF}= \frac{k_x}{\omega}\frac{N^2}{2k_z^2}s^2\begin{bmatrix}
  A^2 \\
  0 \\
  0
\end{bmatrix}={u}_{DF}\begin{bmatrix}
  1 \\
  0 \\
  0
\end{bmatrix},\label{eq:ubf}
\end{eqnarray}
that is, the Bretherton flow is a uniform flow in the $x$-direction with amplitude given by the pseudo-momentum per unit mass.

\bibliographystyle{jfm}
\bibliography{jfm}

\end{document}